\documentclass[journal=jacsat,manuscript=article]{achemso}

\setkeys{acs}{
abbreviations=true,
articletitle=true,
biblabel=plain,
chaptertitle=true,
doi=true,
email=true,
etalmode=truncate,
keywords=true,
maxauthors=100,
super=true
}

\setcitestyle{super,open={},close={}}

\usepackage[utf8]{inputenc}
\usepackage[T1]{fontenc}
\usepackage[english]{babel}
\usepackage[version=4]{mhchem}
\usepackage{siunitx}
\usepackage{graphicx}
\usepackage{tikz}
\usepackage{longtable}
\usepackage{geometry}
\usepackage{subcaption}
\usepackage{float}
\usepackage{setspace}
\usepackage{xkeyval}
\usepackage{array}
\usepackage{booktabs}
\usepackage{listings}
\usepackage{microtype}
\usepackage{hyperref}
\hypersetup{breaklinks,colorlinks=true,linkcolor=black,filecolor=black,urlcolor=black,citecolor=black}

\usepackage{amsmath}
\usepackage{amssymb}
\usepackage{amsfonts}
\usepackage{latexsym}
\usepackage[usenames,dvipsnames]{xcolor}
\usepackage{calc}
\usepackage{multirow}
\usepackage{amstext}
\usepackage{tabularx}
\usepackage{epstopdf}
\usepackage{adjustbox}
\usepackage[section]{placeins}
\usepackage{natmove}
\usepackage{tablefootnote}
\usepackage{threeparttable}
\usepackage[d]{esvect}
\usepackage{mathtools}
\usepackage{physics}
\usepackage{textcomp}
\usepackage[none]{hyphenat}
\usepackage{xr-hyper}
\vfuzz \hfuzz
\SectionsOn
\SectionNumbersOn
\AbstractOn

\title{Transition-Metal Chalcogenide Nanowires: A First-Principles Study of Stability, Electronic Structure, and Mechanical Response}

\author{Danilo de Araújo Moura}
\author{Flavio Bento de Oliveira}
\author{Andreia Luisa da Rosa}
\email{andreialuisa@ufg.br}
\affiliation{Quantum  Computational Materials Science Group, Institute of Physics, Federal University of Goiás, Campus Samambaia, 74690-900, Goiás, Brazil}


\begin{document}

\begin{abstract}
Subnanometer transition-metal chalcogenide nanowires provide a one-dimensional platform in which electronic functionality and mechanical response can be tuned through chemical composition. Here, density functional theory calculations are combined with machine-learning force-field (MLFF) molecular dynamics to investigate MoS, MoSe, MoTe, WS, WSe, and WTe nanowires. The optimized structures are dynamically stable within the harmonic approximation, and their electronic character evolves systematically across the chalcogen series: MoS, MoSe, WS, and WSe are metallic, whereas MoTe and WTe are semiconducting, with HSE06 band gaps of 0.52 and 0.20 eV, respectively. The effective axial stiffness decreases from sulfides to selenides and tellurides, consistent with the increase in metal--chalcogen bond length; using a circular effective cross section, the corresponding Young's moduli range from 392 to 718 GPa. All six nanowires have negative formation energies relative to the selected elemental reference phases. MLFF trajectories extending to 2.5 ns at 300 K preserve the one-dimensional morphology and local atomic order, providing evidence of finite-temperature dynamical stability on the simulated time scale. The corresponding MLFF molecular-dynamics energy traces are shown in. We further examine the response of the nanowires to axial strain and curvature, including strain-induced band-gap modulation in the tellurides. These results establish composition and mechanical deformation as complementary routes for tuning the electronic and mechanical properties of ultrathin transition-metal chalcogenide nanowires and support their further investigation as components of flexible nanoelectronic systems.
\end{abstract}

\maketitle

\section{Introduction}

One-dimensional (1D) nanostructures have attracted considerable attention due to their unique physical properties and potential applications in next-generation electronic, optoelectronic, sensing, and energy-related technologies \cite{Dume2022,Lie2001,Lie2004}. In particular, nanowires exhibit strong quantum confinement, high surface-to-volume ratios, and reduced dimensionality, which often lead to electronic and mechanical properties that differ substantially from their bulk and two-dimensional counterparts \cite{Popov2007,Lin2014}. These characteristics make nanowires promising building blocks for nanoscale devices, including interconnects, transistors, photodetectors, flexible electronics, and wearable sensing platforms \cite{Kim2023,Che2023}.

Among the various classes of low-dimensional materials, transition-metal chalcogenides (TMCs) have emerged as an important family owing to their chemical stability, structural diversity, low toxicity, and highly tunable electronic properties \cite{Wu2021,Wang2022,Yun2020}. While two-dimensional transition-metal dichalcogenides such as MoS$_2$, WS$_2$, MoSe$_2$, and WSe$_2$ have been extensively investigated because of their remarkable electronic and optical properties, significantly less attention has been devoted to their one-dimensional counterparts \cite{Sou2019,Deu2023}. The reduced dimensionality of nanowires may induce substantial modifications in orbital hybridization, bonding characteristics, electronic transport, and mechanical response, leading to properties that are not directly predictable from the corresponding layered materials \cite{Vilfan2006,Popov2007}. 

Recent experimental and theoretical studies have demonstrated the feasibility of synthesizing ultrathin transition-metal chalcogenide nanowires with diameters approaching the subnanometer regime \cite{Lin2014,Popov2007,Deu2023}. These structures exhibit remarkable mechanical robustness and can display either metallic or semiconducting behavior depending on their chemical composition and atomic arrangement \cite{Sou2019,Deu2023}. Such characteristics make them attractive candidates for future nanoelectronic architectures, where mechanical flexibility and electrical functionality must coexist. However, despite these advances, a systematic understanding of the relationship between composition, electronic structure, vibrational stability, and mechanical performance across entire families of transition-metal chalcogenide nanowires remains lacking.

From a technological perspective, identifying nanowires that simultaneously possess high mechanical resilience and favorable electronic properties is particularly important for emerging flexible and wearable electronic devices \cite{Chen2020,He2018,Zhang2019,Gao2019,Gao2020,Yun2019}. In such applications, conductive or semiconducting nanowires can act as active channels, nanoscale interconnects, or functional sensing elements. Nanowire-based systems have already demonstrated promising performance in sensing, energy harvesting, and piezoelectric applications, including ZnO, GaN, and transition-metal sulfide nanowires \cite{Zlw2006,Jol2012,Joa2018,Ros2019,Rod2019}. Therefore, a detailed characterization of their structural stability, elastic behavior, and electronic properties constitutes a necessary first step toward evaluating their suitability for practical device integration. 

In this work, we employ first-principles calculations based on density functional theory to investigate a family of transition-metal chalcogenide nanowires composed of molybdenum and tungsten combined with sulfur, selenium, and tellurium. Specifically, we study MoS, MoSe, MoTe, WS, WSe, and WTe nanowires. We systematically analyze their equilibrium geometries, electronic structures, vibrational properties, charge distributions, mechanical response under strain, and energetic stability relative to elemental reference phases. Our results reveal clear chemical trends across the series, including the emergence of semiconducting behavior in the telluride nanowires and a systematic composition dependence of the axial stiffness. These findings provide a comprehensive understanding of this family of one-dimensional materials and establish a foundation for their future exploration in flexible nanoelectronic applications. While \textit{ab initio} calculations are the gold standard for predicting materials properties, performing long molecular dynamics runs at this level of theory is computationally prohibitive for complex or extended systems. To bridge this gap, we adopted a machine-learning force field (MLFF) framework. This methodology effectively bypasses the immense computational cost of evaluating quantum-mechanical forces at every simulation step, enabling us to investigate the dynamical stability of these nanowires over extended periods without compromising the underlying physical accuracy.

\section{Computational Details}

First-principles calculations were performed within the framework of density functional theory (DFT) using the Vienna \textit{Ab initio} Simulation Package (VASP) \cite{Kresse1999}. The interaction between valence electrons and ionic cores was described using the projector augmented-wave (PAW) method \cite{Kresse1999}, while exchange and correlation effects were treated within the generalized gradient approximation (GGA) using the Perdew--Burke--Ernzerhof (PBE) functional \cite{Perdew1996}. The Kohn--Sham wave functions were expanded in a plane-wave basis set with a kinetic-energy cutoff of 400 eV, and \textit{ab initio} molecular dynamics (AIMD) simulations were performed in the NVT ensemble using a Langevin thermostat to generate reference data for the machine-learning force fields (MLFFs). The reference trajectories were generated at 300 K and sample configurations close to the equilibrium structures.

The Brillouin zone was sampled using a $\Gamma$-centered Monkhorst--Pack mesh of $1\times1\times10$ k-points \cite{Monkhorst1976}. A vacuum region of at least 15 \AA\ was introduced in the directions perpendicular to the nanowire axis to eliminate spurious interactions between periodic images. Spin--orbit coupling (SOC) was included in all electronic-structure calculations owing to its potential importance in transition-metal chalcogenide systems containing heavy elements such as W and Te.

The atomic structures of the MoS, MoSe, MoTe, WS, WSe, and WTe nanowires were fully optimized until the residual forces on all atoms were smaller than $10^{-2}$ eV/\AA. To determine the equilibrium lattice parameter along the wire axis, a series of total-energy calculations was performed for different cell lengths, followed by full relaxation of the atomic positions. The equilibrium geometry was obtained from the minimum of the resulting energy--lattice parameter curve. Periodic DFT calculations of equilibrium, electronic, vibrational, and axial-strain properties employed supercells containing four primitive unit cells (48 atoms). The finite-wire curvature calculations use separate, length-dependent atomistic models described in the bending section.

To assess the dynamical stability of the nanowires, phonon dispersions were calculated within density-functional perturbation theory (DFPT) using the PHONOPY package \cite{PhonopyVASP}. The absence of imaginary phonon frequencies was used as the criterion for dynamical stability.

The electronic properties were analyzed through band structures, total density of states (DOS), projected density of states (PDOS), and charge-density distributions. In addition to PBE calculations, the electronic band gaps of the semiconducting systems were refined using the screened hybrid functional HSE06, which incorporates 25\% exact exchange in the short-range part of the screened hybrid functional. Hartree--Fock exchange and a screening parameter of 0.2 \AA$^{-1}$. The HSE06 calculations were performed using the PBE-optimized geometries.

Mechanical properties were investigated through uniaxial deformations applied along the nanowire axis. Strain was defined as

\begin{equation}
\varepsilon=\frac{L-L_0}{L_0},
\end{equation}

where $L_0$ and $L$ are the equilibrium and strained supercell lengths, respectively. Both tensile and compressive strains were applied in increments of 1\% within the elastic regime, and all atomic positions were fully relaxed for each strain value.

The stress was obtained from the derivative of the total energy with respect to strain,

\begin{equation}
\sigma=\frac{1}{V}\frac{\partial E}{\partial \varepsilon},
\end{equation}

where $V$ is the effective volume of the nanowire. Following previous studies on one-dimensional nanostructures \cite{Sou2019}, the nanowire cross-sectional area was approximated as a circular section with radius equal to the average distance of the outermost chalcogen atoms from the wire axis. Because a unique cross-sectional area cannot be defined for an atomically thin one-dimensional object, the resulting Young's moduli should be interpreted as effective values associated with the adopted circular cross-section convention. The Young's modulus was determined from the slope of the linear region of the stress--strain curve, while the axial elastic constant was obtained from

\begin{equation}
k=\frac{YA}{L},
\end{equation}

where $Y$ is the Young's modulus, $A$ is the effective cross-sectional area, and $L$ is the nanowire length.

The energetic stability of the nanowires was evaluated through formation energies relative to selected elemental reference phases. A negative value indicates that formation of the nanowire is energetically favorable relative to those elemental references. This criterion does not, by itself, establish stability against all competing binary or multinary phases.

\section{Results and Discussion}
\subsection{Structural and Electronic Properties}

\begin{figure}[htbp]
\centering
\includegraphics[width=\linewidth,trim=-5mm -3mm -5mm -3mm,clip]{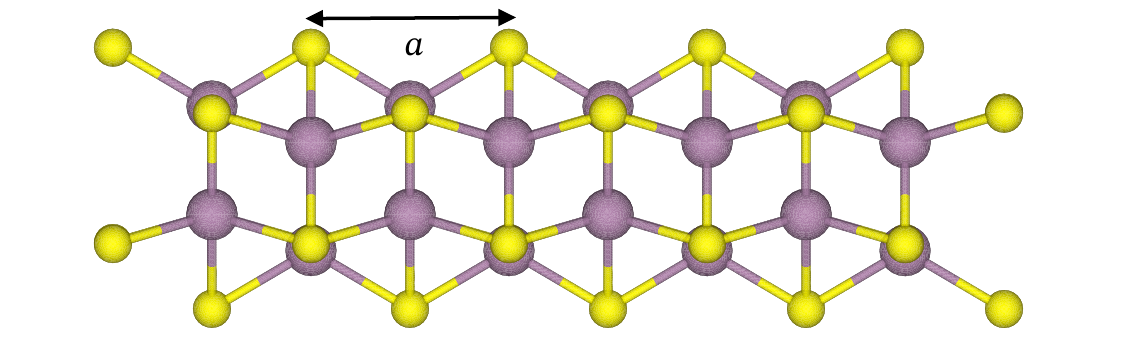}
\caption{Cross-sectional views of MX (M = Mo,W; X = S,Se,Te) nanowires.}
\label{fig:structures_main}
\end{figure}

The schematic cross-sectional structures of the nanowires is shown in Fig.~\ref{fig:structures_main}. Complete side views are provided in the Supporting Information. All investigated systems preserve the characteristic staggered-triangle geometry previously reported for subnanometer transition-metal chalcogenide nanowires \cite{Vilfan2006,Cakir2006,Popov2007,Sou2019,Deu2023}. The calculated lattice parameter of the MoS nanowire ($a=4.34$~\AA) is in excellent agreement with previous theoretical and experimental investigations \cite{Lin2014,Sou2019,Deu2023}. Similar agreement is observed for the corresponding bond lengths, confirming the reliability of the adopted computational approach.

A systematic increase in the lattice parameter and metal--chalcogen bond lengths is observed when moving from sulfur to selenium and tellurium compounds (Table~\ref{tab:structural_electronic}). This behavior originates from the larger atomic radii of the heavier chalcogen atoms and reflects the progressive weakening of the metal--chalcogen bonds along the series. In contrast, replacing Mo by W produces only minor structural modifications, indicating that the overall geometry is primarily controlled by the chalcogen species.

\begin{figure}[htbp]
\centering
\includegraphics[width=\linewidth,clip]{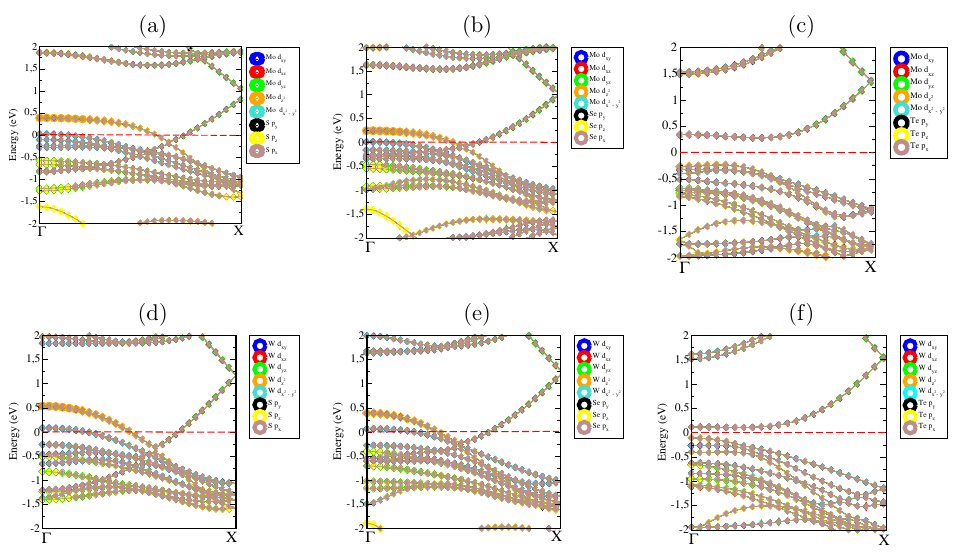}
\caption{Projected equilibrium electronic band structures of (a) MoS, (b) MoSe, (c) MoTe, (d) WS, (e) WSe, and (f) WTe. MoS, MoSe, WS, and WSe are metallic, whereas MoTe and WTe are semiconducting.}
\label{fig:bands_main}
\end{figure}

The projected equilibrium electronic band structures are summarized in Fig.~\ref{fig:bands_main} and reveal distinct trends across the investigated family. The MoS, MoSe, WS, and WSe nanowires exhibit metallic behavior, characterized by bands crossing the Fermi level, whereas MoTe and WTe display semiconducting character with finite band gaps (Table~\ref{tab:structural_electronic}). The emergence of semiconducting behavior in the telluride nanowires can be attributed to the stronger participation of Te $5p$ states and the corresponding modification of the hybridization between transition-metal $d$ orbitals and chalcogen $p$ orbitals. Similar dimensionality-induced electronic transitions have been predicted previously for transition-metal chalcogenide nanostructures \cite{Popov2007,Sou2019,Deu2023}.

The phonon dispersions provided in Fig.~S1 of the Supporting Information further demonstrate the dynamical stability of all investigated nanowires. No imaginary phonon frequencies are observed throughout the Brillouin zone, indicating that the optimized structures correspond to stable local minima of the potential-energy surface. The low-frequency region exhibits the characteristic vibrational signatures of quasi-one-dimensional systems, including one longitudinal acoustic mode, one torsional acoustic mode, and two flexural acoustic modes displaying quadratic dispersion near the $\Gamma$ point. Such behavior is consistent with previous theoretical analyses of ultrathin transition-metal chalcogenide nanowires \cite{Vilfan2006,Popov2007}.

The electronic origin of the metallic and semiconducting behavior can be understood from the orbital-resolved DOS and PDOS provided in the Supporting Information. The corresponding orbital-resolved DOS and PDOS are provided in Figs.~S2--S3 of the Supporting Information. For MoS, MoSe, WS, and WSe, the states near the Fermi level are dominated by transition-metal $d$ orbitals, particularly the $d_{xy}$, $d_{x^{2}-y^{2}}$, and $d_{z^{2}}$ components. The strong contribution of these orbitals leads to partially occupied bands crossing the Fermi energy, thereby producing metallic conductivity. In contrast, the telluride nanowires exhibit enhanced hybridization between transition-metal $d$ states and Te $p$ orbitals, which results in the opening of a finite band gap.

The charge-density distributions reported in the Supporting Information provide additional insight into the bonding characteristics. In all nanowires, the electronic density is concentrated along the metal--chalcogen bonds, indicating a predominantly covalent bonding nature. However, the spatial distribution becomes progressively more diffuse when moving from sulfides to tellurides, reflecting the larger spatial extension of the chalcogen orbitals and the weaker bond strengths associated with heavier chalcogen atoms. This trend correlates directly with the observed reduction in mechanical stiffness discussed below.

\begin{table*}[t]
\centering
\caption{Optimized structural parameters and electronic properties of the
transition-metal chalcogenide nanowires. The lattice parameter $a$ is measured
along the wire axis. Bond lengths are grouped according to the inner and outer
parts of the nanowire. A dash indicates metallic behavior.}
\label{tab:structural_electronic}
\small
\setlength{\tabcolsep}{5.5pt}
\renewcommand{\arraystretch}{1.18}

\begin{tabular}{lcccccclcc}
\toprule
& \multicolumn{5}{c}{\text{Structural properties}}
& \multicolumn{3}{c}{\text{Electronic properties}} \\
\cmidrule(lr){2-6}
\cmidrule(lr){7-9}

\text{Nanowire}
& $\boldsymbol{a}$ (\AA)
& \multicolumn{2}{c}{\text{Inner bonds (\AA)}}
& \multicolumn{2}{c}{\text{Outer bonds (\AA)}}
& \text{Character}
& \multicolumn{2}{c}{\text{Band gap (eV)}} \\

\cmidrule(lr){3-4}
\cmidrule(lr){5-6}
\cmidrule(lr){8-9}

&
& M--X & M--M
& M--X & M--M
&
& PBE & HSE06 \\

\midrule

MoS
& 4.34
& 2.48 & 2.74
& 2.52 & 2.68
& Metallic
& -- & -- \\

MoSe
& 4.43
& 2.61 & 2.72
& 2.65 & 2.71
& Metallic
& -- & -- \\

MoTe
& 4.58
& 2.81 & 2.65
& 2.86 & 2.75
& Semiconducting
& 0.2474 & 0.5155 \\

\addlinespace[2pt]

WS
& 4.36
& 2.50 & 2.73
& 2.54 & 2.69
& Metallic
& -- & -- \\

WSe
& 4.43
& 2.64 & 2.72
& 2.66 & 2.71
& Metallic
& -- & -- \\

WTe
& 4.54
& 2.85 & 2.66
& 2.86 & 2.74
& Semiconducting
& 0.0890 & 0.1977 \\

\bottomrule
\end{tabular}
\end{table*}

\subsection{Mechanical Properties}

\begin{figure}[htbp]
\centering
\includegraphics[width=\linewidth,clip]{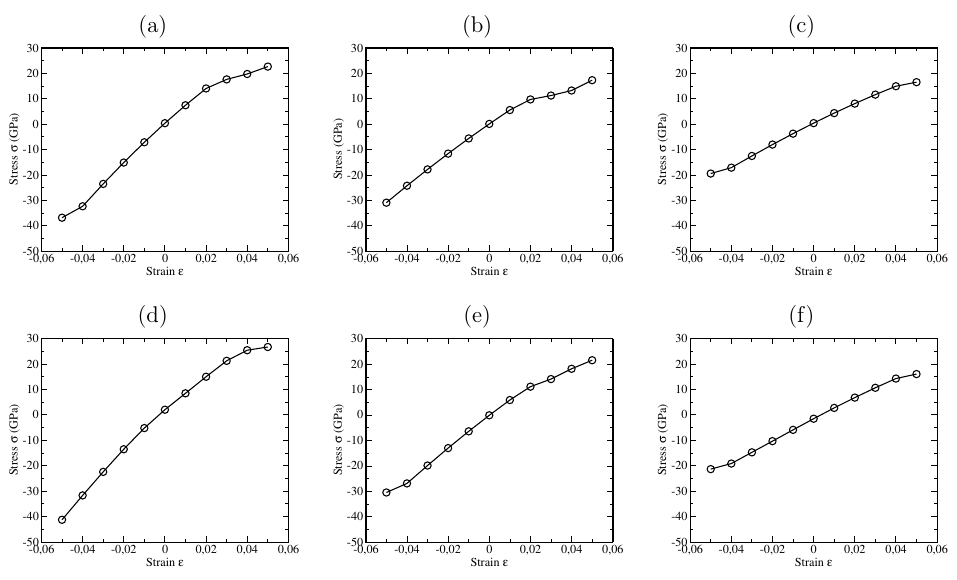}
\caption{Stress--strain response in the elastic regime for \text{(a)} MoS, \text{(b)} MoSe, \text{(c)} MoTe, \text{(d)} WS, \text{(e)} WSe, and \text{(f)} WTe. The slopes near zero strain yield the effective Young's moduli summarized in Table~\ref{tab:mechanical_properties}.}
\label{fig:stress_strain_main}
\end{figure}

The mechanical response of the transition-metal chalcogenide nanowires was investigated through uniaxial strain calculations within the elastic regime. The near-equilibrium stress--strain relationships are summarized in Fig.~\ref{fig:stress_strain_main}, while the corresponding energy--strain curves and the complete near-equilibrium and extended mechanical datasets are provided in Figs.~S16--S22 of the Supporting Information. Representative structural evolution up to large deformation and rupture is shown in Figs.~S4--S9 of the Supporting Information. In all systems, the total energy exhibits a nearly parabolic dependence on strain near equilibrium, confirming that the applied deformations remain within the harmonic regime and allowing a reliable determination of the elastic properties.

The calculated Young's moduli are summarized in Table~\ref{tab:mechanical_properties}. Among the investigated systems, the sulfide nanowires exhibit the largest stiffness, with values of approximately 705~GPa and 718~GPa for MoS and WS, respectively. These values are remarkably large for one-dimensional materials and are comparable to those reported for other mechanically robust nanowire systems \cite{Vilfan2006,Sou2019}. The large Young's moduli originate from the strong covalent character of the metal--sulfur bonds and the compact geometry of the nanowires.

A clear chemical trend is observed across both the Mo- and W-based series. The Young's modulus decreases systematically from sulfides to selenides and tellurides. For example, the modulus decreases from 705~GPa (MoS) to 533~GPa (MoSe) and further to 392~GPa (MoTe). An analogous trend is observed for the tungsten compounds. This behavior can be directly associated with the increase in bond length and the reduction in bond strength as the chalcogen atomic size increases.

The elastic constants follow the same trend, decreasing from approximately 120~N/m for sulfides to about 80~N/m for tellurides. Despite this reduction, all investigated nanowires exhibit substantial axial stiffness. The large effective Young's moduli quantify resistance to axial stretching and should not be interpreted as a direct measure of bending flexibility; the latter is assessed separately through the curvature-dependent calculations discussed below. The combination of reduced dimensionality, structural integrity, and composition-dependent electronic properties motivates further evaluation of these nanowires for nanoscale interconnects and flexible nanoelectronic components.

\begin{table}[t]
\centering
\caption{Mechanical properties of the Mo- and W-based nanowires. The effective cross-sectional area $A$ is used to obtain Young's modulus $Y$, while $k$ denotes the axial elastic constant.}
\label{tab:mechanical_properties}
\small
\renewcommand{\arraystretch}{1.15}
\begin{tabular}{lccc}
\toprule
\text{Nanowire} & $\boldsymbol{A}$ (\AA$^2$) & $\boldsymbol{Y}$ (GPa) & $\boldsymbol{k}$ (N m$^{-1}$) \\
\midrule
MoS  & 29 & 705 & 119 \\
MoSe & 32 & 533 &  97 \\
MoTe & 37 & 392 &  79 \\
\addlinespace[2pt]
WS   & 30 & 718 & 123 \\
WSe  & 33 & 552 & 103 \\
WTe  & 38 & 397 &  83 \\
\bottomrule
\end{tabular}
\end{table}

\subsection{Strain-Induced Electronic Response and Large-Strain Evolution}

The response to axial deformation was examined at two distinct levels. Small tensile and compressive strains were used to determine elastic properties from the near-equilibrium energy--strain and stress--strain relations. Larger tensile strains were considered separately to visualize structural rearrangements and the approach to rupture. The latter regime is intrinsically nonlinear and is therefore not used to extract Young's moduli.

\begin{figure}[htbp]
\centering
\includegraphics[width=\linewidth,clip]{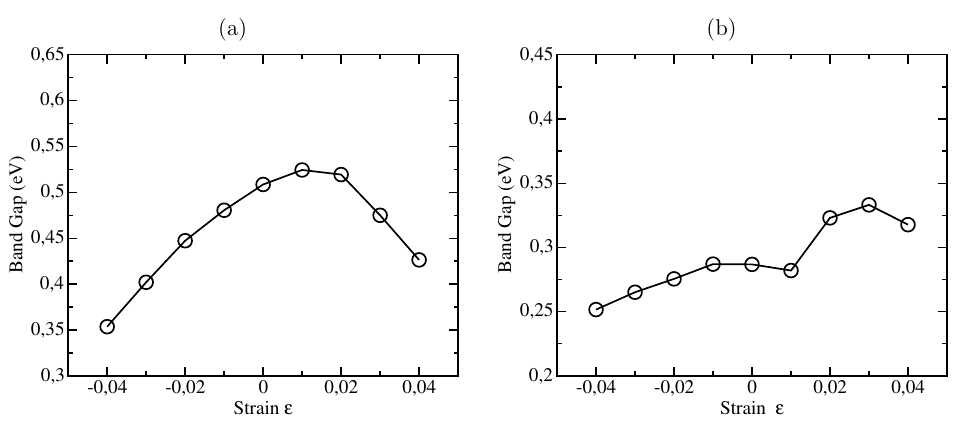}
\caption{HSE06 band gap as a function of axial strain for \text{(a)} MoTe and \text{(b)} WTe over the interval $-4\%\leq\varepsilon\leq+4\%$. The strain dependence demonstrates direct electromechanical tuning of the semiconducting telluride nanowires.}
\label{fig:gap_strain_hse}
\end{figure}

The strain-dependent band structures show that the metallic nanowires remain metallic over the small-strain interval considered, while the semiconducting tellurides retain finite gaps whose magnitudes vary with strain. The complete strain-dependent band structures are provided in Figs.~S10--S15 of the Supporting Information. The HSE06 results in Figure~\ref{fig:gap_strain_hse} therefore demonstrate a direct coupling between axial deformation and the electronic structure of MoTe and WTe. This strain tunability is more directly relevant to prospective electromechanical functionality than the magnitude of the axial modulus alone.

The large-strain structural sequences are reported in the Supporting Information. They should be interpreted as quasistatic pathways toward rupture rather than as extensions of the linear elastic regime, because the associated configurations involve substantial bond rearrangement and nonlinear response.

The complete PBE band structures for strains from $-4\%$ to $+4\%$ for all six nanowires are reported in the Supporting Information. In the main text we focus on the quantitatively refined HSE06 gap response of the semiconducting tellurides.

Because the effective cross-sectional area of an atomically thin one-dimensional object is not uniquely defined, the Young's moduli in Table~\ref{tab:mechanical_properties} should be interpreted using the stated circular-area convention. The corresponding axial elastic constant, $k=YA/L$, provides a complementary measure of resistance to axial deformation.

\section{Formation Energy Relative to Elemental References}

The energetic favorability of the investigated nanowires relative to their elemental reference phases was evaluated through their formation energies.
The formation energy was obtained from

\begin{equation}
E_f =
E_{tot}^{NW}
-
n_M \mu_M
-
n_X \mu_X ,
\end{equation}

where $E_{tot}^{NW}$ is the total energy of the nanowire,
$n_M$ and $n_X$ are the numbers of metal and chalcogen atoms,
and $\mu_M$ and $\mu_X$ correspond to the energies per atom
of the elemental reference phases.

The calculated formation energies are summarized in Table~\ref{tab:formation}.

The calculated formation energy for the MoS$_2$ sheet is 
$\Delta H_{\mathrm{MoS_2}} = -2.57 $, 
in good agreement with the experimental value of $-2.44$ eV at room temperature. For the pristine MoS nanowire, the computed formation energy is $\Delta_{MoS} = -1.05 \ \mathrm{eV}$.

The negative value shows that formation of the MoS nanowire is energetically favorable relative to the selected elemental reference phases. The same criterion gives negative values for MoSe, MoTe, WS, WSe, and WTe. These results should not be interpreted as a complete thermodynamic-stability analysis against all competing phases; such a statement would require a chemical-potential or convex-hull treatment.\\

\begin{table}[t]
\centering
\caption{Calculated formation energies of the investigated nanowires and
the corresponding layered transition-metal dichalcogenides. More negative
values indicate greater stability with respect to decomposition into the
elemental reference phases.}
\label{tab:formation}
\small
\setlength{\tabcolsep}{8pt}
\renewcommand{\arraystretch}{1.15}

\begin{tabular}{lclc}
\toprule
\multicolumn{2}{c}{\text{Nanowire}}
&
\multicolumn{2}{c}{\text{Layered TMD}} \\
\cmidrule(lr){1-2}
\cmidrule(lr){3-4}

\text{Material}
& $\boldsymbol{\Delta H_f}$ (eV/f.u.)
& \text{Material}
& $\boldsymbol{\Delta H_f}$ (eV/f.u.) \\

\midrule

MoS  & $-1.05$ & MoS$_2$  & $-2.57$ \\
MoSe & $-0.87$ & MoSe$_2$ & $-1.99$ \\
MoTe & $-0.43$ & MoTe$_2$ & $-0.75$ \\

\addlinespace[2pt]

WS   & $-0.65$ & WS$_2$   & $-2.35$ \\
WSe  & $-0.48$ & WSe$_2$  & $-1.57$ \\
WTe  & $-0.03$ & WTe$_2$  & $-0.08$ \\

\bottomrule
\end{tabular}
\end{table}

The formation energies of the nanowires are systematically
less negative than those of the corresponding layered
transition-metal dichalcogenides. This behavior is expected
due to the reduced coordination number and lower bonding
connectivity in one-dimensional structures. All investigated nanowires nevertheless exhibit negative formation energies relative to the selected elemental references.

A clear chemical trend is also observed, with the magnitude
of the formation enthalpy decreasing from sulfides to
selenides and tellurides. This trend reflects the progressive
weakening of the metal--chalcogen bonds as the chalcogen
atomic size increases and is consistent with the reduction
in Young's modulus observed across the same series.
\section{Finite-Temperature Stability from MLFF Molecular Dynamics}

Standard \textit{ab initio} molecular dynamics (AIMD) simulations provide highly accurate descriptions of atomic interactions, but their steep computational cost limits the accessible time and length scales. To overcome this bottleneck and efficiently assess the stability of these nanowires, molecular dynamics simulations were performed using a machine-learning force field (MLFF) based on \textit{ab initio} calculations. The MLFF was trained at 300 K over a 20 ps trajectory for each structure. The resulting root-mean-square errors (RMSE) for energy and forces—detailed in Table \ref{tab:bayesianandERRE}—were notably low (in the meV range), accompanied by Bayesian errors on the order of $10^{-5}$~eV/atom. These errors indicate good agreement with the reference data within the near-equilibrium training domain. Because the training trajectories sample configurations close to equilibrium at 300 K, transferability to strongly strained or highly curved configurations requires separate validation.
	\begin{table}[t]
\centering
\caption{Validation errors of the machine-learning force fields trained at 300 K. The Bayesian error quantifies the model uncertainty, whereas RMSE$_E$ and RMSE$_F$ are the root-mean-square errors in energies and atomic forces, respectively.}
\label{tab:bayesianandERRE}
\small
\renewcommand{\arraystretch}{1.15}
\begin{tabular}{lccc}
\toprule
\text{Nanowire} & \text{Bayesian error} & \text{RMSE$_E$} & \text{RMSE$_F$} \\
& (eV atom$^{-1}$) & (meV atom$^{-1}$) & (meV \AA$^{-1}$) \\
\midrule
MoS  & $3.75\times10^{-5}$ & 6.86 & 15.53 \\
MoSe & $2.93\times10^{-5}$ & 2.50 & 75.27 \\
MoTe & $1.40\times10^{-5}$ & 0.48 & 35.60 \\
\addlinespace[2pt]
WS   & $3.53\times10^{-5}$ & 0.78 & 62.40 \\
WSe  & $2.48\times10^{-5}$ & 2.24 & 78.71 \\
WTe  & $6.41\times10^{-6}$ & 0.98 & 16.66 \\
\bottomrule
\end{tabular}
\end{table}

Leveraging the computational efficiency of the MLFFs, production runs were extended to 2.5 ns to probe finite-temperature dynamical stability at 300 K. The energy signals fluctuate around stationary mean values throughout the trajectories (Fig.~S23 of the Supporting Information). This behavior, together with the structural analyses below, indicates that no spontaneous reconstruction or loss of the one-dimensional morphology occurs on the simulated time scale.

\begin{figure}[tb]
\centering
\includegraphics[width=\linewidth,clip]{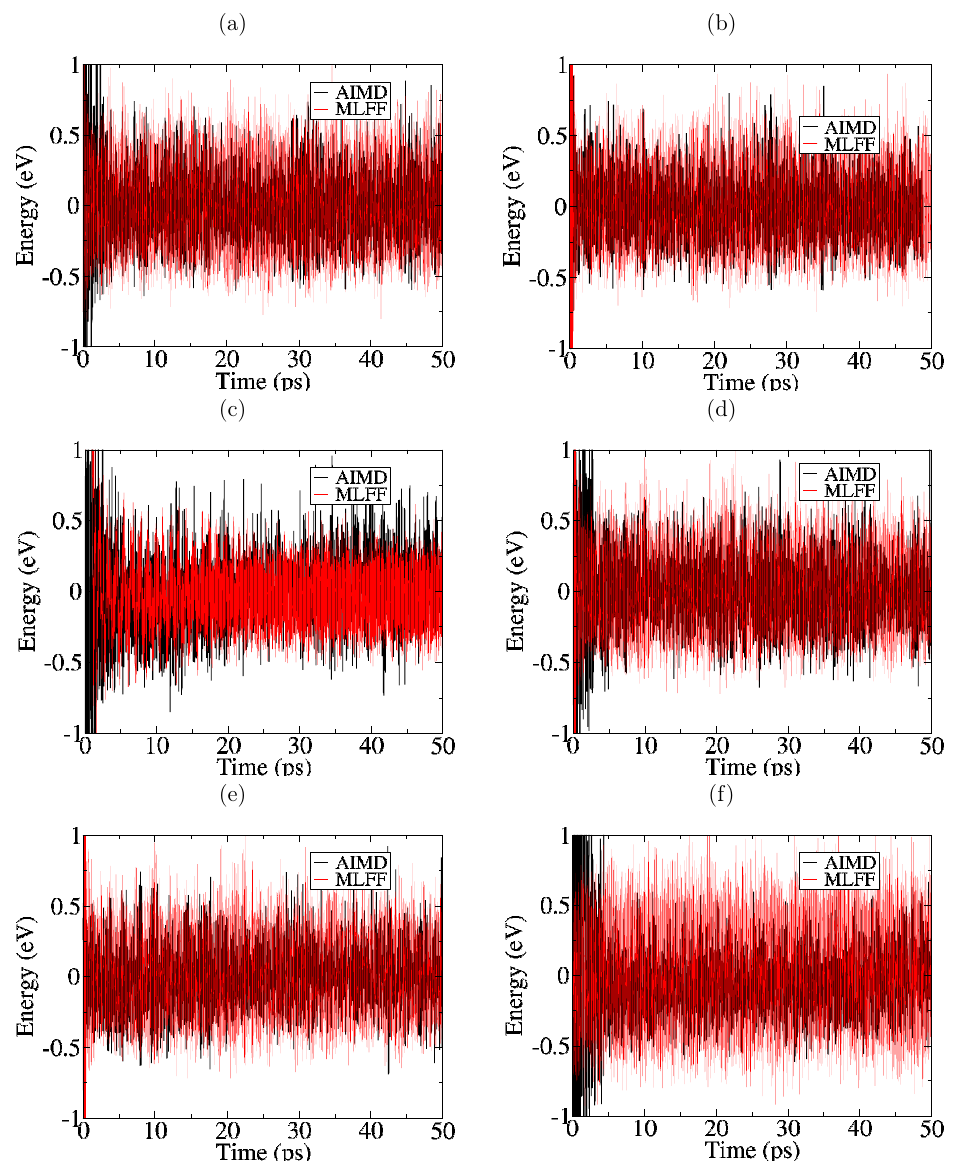}
\caption{Comparison of the total energy fluctuations over time for the investigated nanowires, calculated via Ab Initio Molecular Dynamics (AIMD, black line) and the trained Machine Learning Force Field (MLFF, red line). Both trajectories are zero-centered with respect to their average energies. The agreement in the thermal energy fluctuations over the 50 ps interval supports the accuracy of the MLFF within the near-equilibrium configurational domain sampled at 300 K. Strongly strained or highly curved configurations require separate validation. The panels correspond to the nanowires of \text{(a)} MoS, \text{(b)} MoSe, \text{(c)} MoTe, \text{(d)} WS, \text{(e)} WSe, and \text{(f)} WTe.}
\label{fig:validation_nanowires}
\end{figure}

To assess the reliability of the MLFFs within the thermally sampled near-equilibrium regime, the trained models were compared with first-principles trajectories. Figure~\ref{fig:validation_nanowires} displays the total-energy fluctuations of the investigated nanowires over a 50~ps interval, calculated using both ab initio molecular dynamics (AIMD) and the MLFF. The agreement in the mean-subtracted energy fluctuations indicates that the MLFFs reproduce the DFT energetics well for configurations sampled close to equilibrium at 300 K. This comparison should not, however, be interpreted as a validation of the force field over the entire potential-energy surface, particularly for strongly strained, highly curved, buckled, or locally reconstructed configurations.

\begin{figure}[htbp]
\centering
\includegraphics[width=\linewidth,clip]{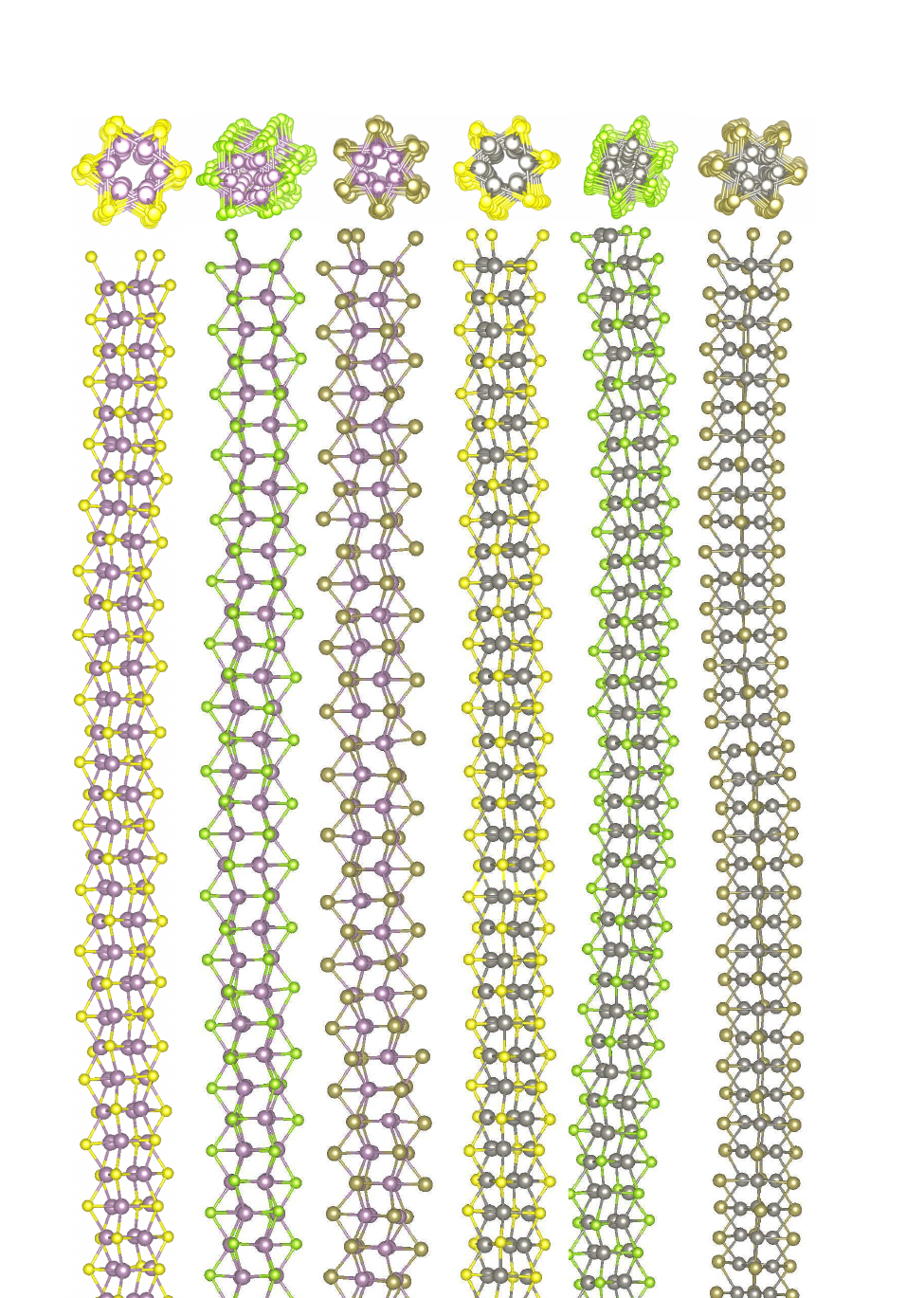}
\caption{Final structural snapshots of the investigated nanowires after 2.5 ns of molecular dynamics simulations at 300 K using the machine-learning force field (MLFF). From left to right, the snapshots correspond to MoS, MoSe, MoTe, WS, WSe, and WTe. The preservation of the characteristic one-dimensional morphology indicates structural stability throughout the simulations.}
\label{fig:mlff_snapshots}
\end{figure}

To visually corroborate this stability, final snapshots of the nanowires extracted at the end of the 2.5 ns MD runs are presented in Figure~\ref{fig:mlff_snapshots}. Both the lateral and cross-sectional views demonstrate that the one-dimensional morphology and the characteristic inner pore structures remain fully intact, without any signs of bond breaking or significant structural distortion.

\begin{figure}[tb]
\centering
\includegraphics[width=\linewidth,clip]{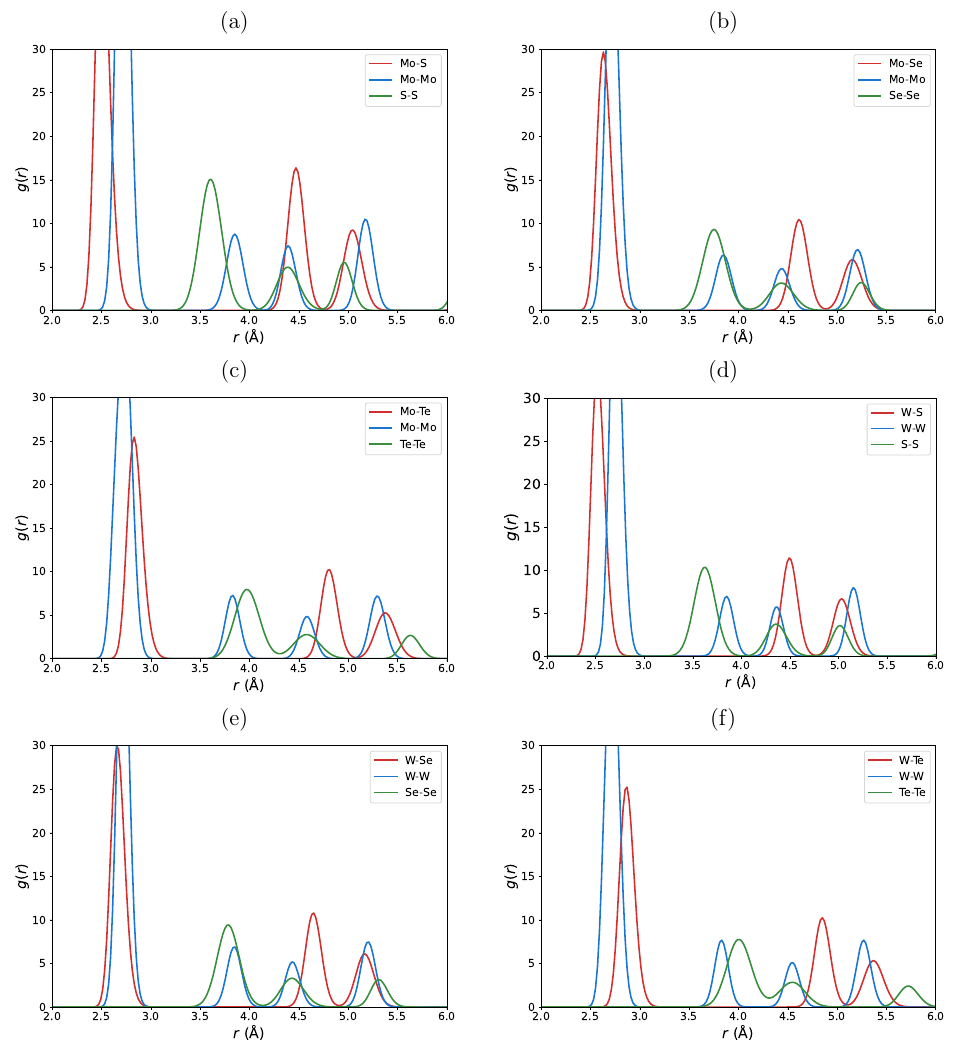}
\caption{Radial distribution function (RDF) profiles calculated from
the 2.5 ns MLFF molecular dynamics trajectories at 300 K. The sharp,
well-defined peaks for the metal--chalcogen, metal--metal, and
chalcogen--chalcogen pairs are consistent with preservation of local
atomic order and show no indication of amorphization over the simulated
time interval. The panels correspond to the nanowires of
\text{(a)} MoS, \text{(b)} MoSe, \text{(c)} MoTe,
\text{(d)} WS, \text{(e)} WSe, and \text{(f)} WTe.}

\label{fig:RDF_nanowires}
\end{figure}

Furthermore, to quantitatively verify that the wires maintain this structural integrity over time, we analyzed the radial distribution function (RDF), which provides a robust measure of local atomic ordering. The resulting profiles, presented in Figure~\ref{fig:RDF_nanowires}, exhibit sharp, well-defined coordination peaks for the respective metal-chalcogen (M-X), metal-metal (M-M), and chalcogen-chalcogen (X-X) pairs (where M = Mo, W and X = S, Se, Te), which is consistent with Table~\ref{tab:structural_electronic} regarding the first peaks of each nanowire. The preservation of these distinct crystalline peaks after 2.5 ns is in excellent agreement with the visual snapshots, showing that the nanowires retain local crystalline order without evidence of melting or amorphization over the simulated 2.5 ns interval.

\subsection{Curvature Response, Harmonic Bending, and Structural Instabilities}

The response to curvature was analyzed by comparing the energy of a bent nanowire with that of the corresponding straight configuration. The bending energy is defined as $E_{\mathrm{bend}}(R)=E(R)-E_{\mathrm{straight}}$, where $R$ is the imposed radius of curvature. For a uniformly bent one-dimensional object in the harmonic regime, continuum elasticity predicts

\begin{equation}
E_{\mathrm{bend}} = \frac{1}{2}D\frac{L}{R^2},
\end{equation}

or, equivalently,

\begin{equation}
\frac{E_{\mathrm{bend}}}{L}=\frac{D}{2}\kappa^2,
\qquad \kappa=\frac{1}{R},
\end{equation}

where $L$ is the nanowire length, $\kappa$ is the curvature, and $D$ is the bending rigidity. Previous studies of Mo$_6$S$_6$ have reported high bending flexibility, weak changes in electronic properties under smooth bending, and a richer mechanical response under twisting; more recent work has also examined kinked geometries \cite{Popov2008Bending,Xu2023,KinkedMo6S6_2024}. Consequently, a length-independent intrinsic bending rigidity can only be assigned when $E_{\mathrm{bend}}/L$ varies linearly with $\kappa^2$ and the deformation remains approximately uniform. In the following, values obtained from local linear fits are therefore referred to as \emph{apparent bending rigidities} unless these conditions are satisfied.

The present calculations reveal two distinct mechanical regimes. At sufficiently weak curvature, the nanowires can accommodate deformation predominantly through smooth bending, and the quadratic continuum expression provides a useful local description. At larger curvature and/or increasing wire length, however, additional relaxation channels become accessible. The compressed side of the wire can develop nonuniform distortions, including rippling, local buckling, shear-like rearrangements, and departures from a circular arc. Once such modes appear, the deformation is no longer described by a single uniform curvature and the fitted coefficient should not be interpreted as an intrinsic material constant.

This crossover is particularly clear for MoS. The apparent rigidities extracted from the currently available datasets are 268.91, 209.92, 175.29, 158.77, and 89.40~$\mathrm{eV}\,\text{\AA}$ for $L=10$, 25, 50, 100, and 200~nm, respectively. Rather than approaching a length-independent plateau, the fitted value decreases systematically over this range. The longest structures also exhibit visible undulations that relieve compressive strain. The length dependence is therefore interpreted here as evidence that the calculations progressively leave the uniform harmonic-bending regime, rather than as a size dependence of an intrinsic $D$.

The same analysis provides a natural interpretation of the nonmonotonic behavior found for the other compositions. Abrupt reductions in the apparent rigidity, such as those observed for selected MoSe and WS geometries, coincide with the onset of nonuniform structural relaxation. The Se- and Te-containing nanowires show a stronger tendency toward these deformation modes, consistent with their lower axial stiffness and longer metal--chalcogen bonds. The relevant result is thus not a single bending-rigidity number for each material, but the existence of a composition- and geometry-dependent crossover from approximately harmonic bending to a nonlinear regime dominated by structural relaxation.

Because the MLFFs were trained using configurations sampled close to equilibrium at 300 K, the strongly curved configurations probe a broader configurational domain than the training trajectories. Quantitative bending rigidities should therefore be regarded as provisional until representative straight, weakly bent, and strongly bent configurations are benchmarked directly against DFT energies and forces. Such benchmarks will distinguish genuine mechanical softening and instability from possible MLFF extrapolation. The present curvature calculations nevertheless identify the relevant deformation mechanisms and show that axial stiffness alone is insufficient to characterize the mechanical flexibility of these ultrathin nanowires.

\begin{figure}[htbp]
\centering
\includegraphics[width=\linewidth,clip]{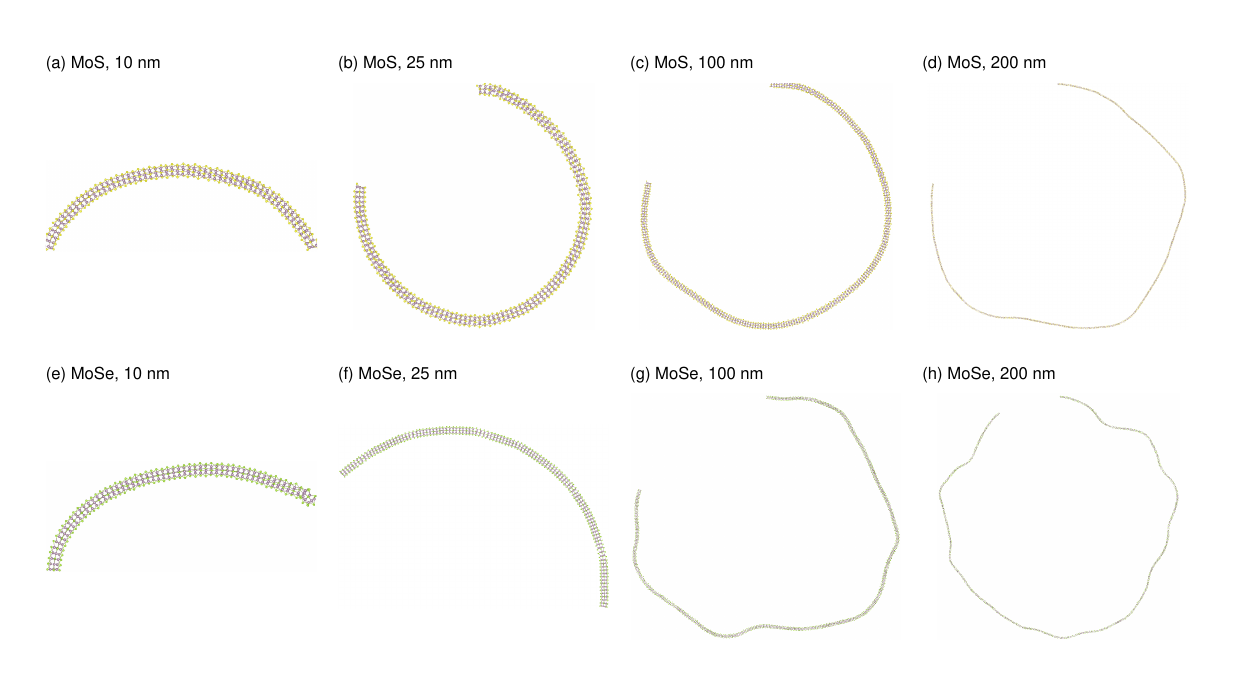}
\caption{Structural snapshots of bent MoS and MoSe nanowires for different wire lengths $L$ and imposed bending configurations. Multiple snapshots at the same $L$ correspond to different imposed bending configurations.}
\label{fig:bent_nanowires_snapshots}
\end{figure}

\begin{figure}[htbp]
\centering
\includegraphics[width=\linewidth,clip]{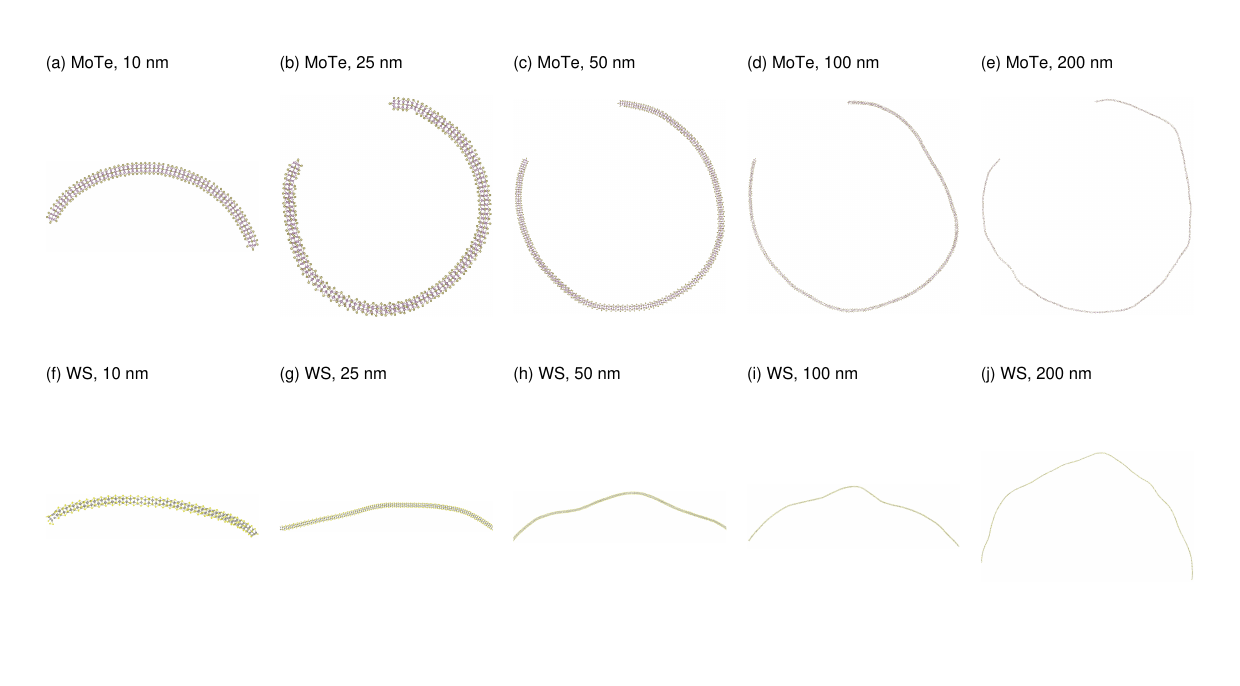}
\caption{Structural snapshots of bent MoTe and WS nanowires for different wire lengths $L$ and imposed bending configurations. Multiple snapshots at the same $L$ correspond to different imposed bending configurations.}
\label{fig:bent_mote_ws}
\end{figure}

\begin{figure}[htbp]
\centering
\includegraphics[width=\linewidth,clip]{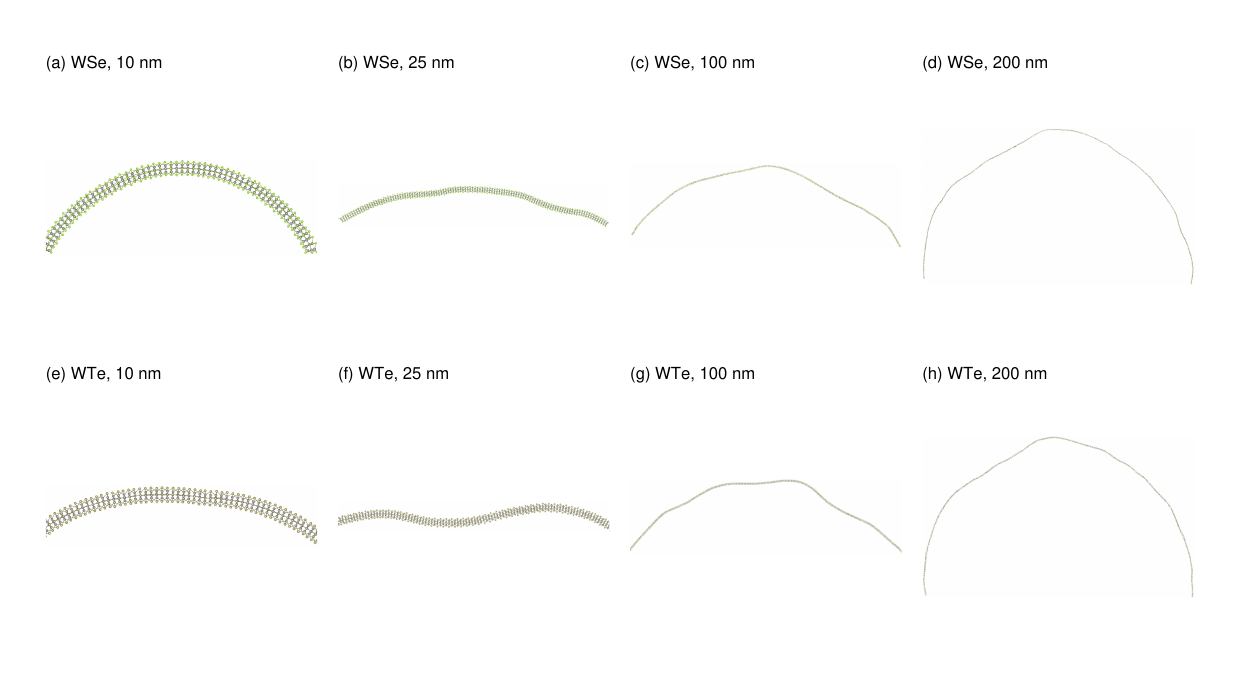}
\caption{Structural snapshots of bent WSe and WTe nanowires for different wire lengths $L$ and imposed bending configurations. Multiple snapshots at the same $L$ correspond to different imposed bending configurations.}
\label{fig:bent_wse_wte}
\end{figure}


\begin{figure}[htpb]
\centering
\includegraphics[width=\linewidth,clip]{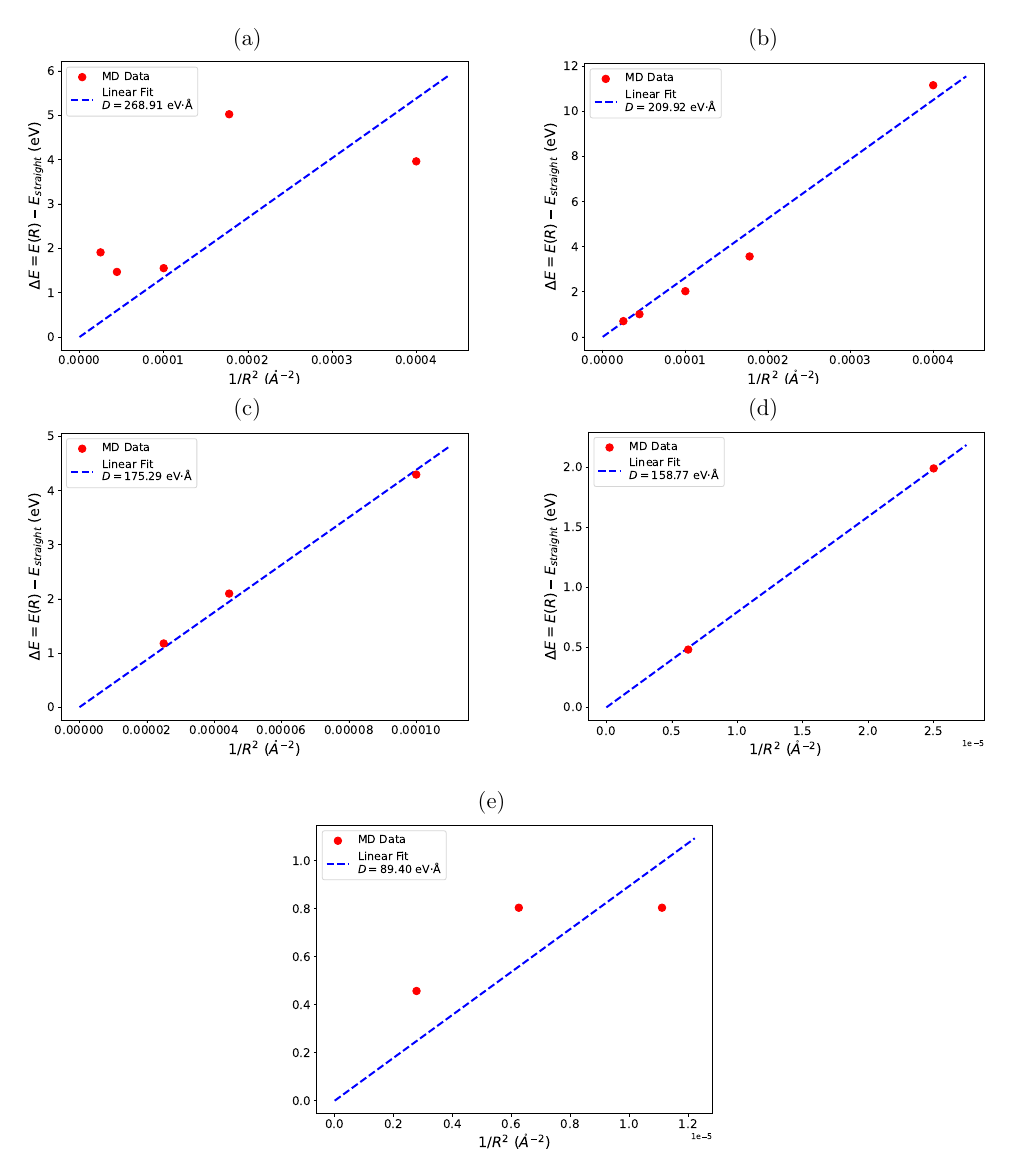}
\caption{Normalized bending energy $\Delta E / L$ as a function of the inverse square of the curvature radius ($1/R^2$) for MoS nanowires of varying lengths: (a) $10$ nm, (b) $25$ nm, (c) $50$ nm, (d) $100$ nm, and (e) $200$ nm. The dashed lines represent local linear fits over the sampled curvature range. The quantity $D_{\mathrm{app}}=2\,d(\Delta E/L)/d(1/R^2)$ is therefore reported as an apparent bending rigidity. Its substantial variation with $L$ shows that a unique length-independent continuum bending constant has not been established by the present dataset.}
\label{fig:bending_lengths}
\end{figure}

\begin{table}[htpb]
	\centering
	\caption{Apparent bending rigidity $D_{\mathrm{app}}$ (in $\mathrm{eV}\,\text{\AA}$) obtained from local linear fits of $\Delta E/L$ versus $1/R^2$ for the sampled nanowire lengths. The pronounced length and composition dependence indicates that these fitted values should not be interpreted as intrinsic continuum constants. Configurations showing nonuniform deformation must be assessed structurally and benchmarked against DFT before a quantitative intrinsic rigidity is assigned.}
	\label{tab:bending_rigidity}
	\begin{tabular}{lccccc}
		\hline\hline
		Nanowire & $L = 10$ nm & $L = 25$ nm & $L = 50$ nm & $L = 100$ nm & $L = 200$ nm \\ \hline
		MoS      & 268.91      & 209.92      & 175.29      & 158.77       & 89.40        \\ \hline
		MoSe     & 391.10      & 23.00      & 417.62      & 267.91       & 376.44       \\ \hline
		MoTe     & 207.82      & 112.66      & 166.62      & 166.59       & 91.61        \\ \hline
		WS       & 467.18      & 214.17     & 431.26      & 270.58       & 185.66       \\ \hline
		WSe      & 13.31       & 26.85       & 56.08       & 106.87       & 211.02       \\ \hline
		WTe      & 17.09       & 153.31      & 58.29       & 82.75        & 130.96       \\ \hline\hline
	\end{tabular}
\end{table}

\section{Conclusion}

We have systematically investigated the structural, electronic, vibrational, energetic, and mechanical properties of MoS, MoSe, MoTe, WS, WSe, and WTe one-dimensional nanowires using first-principles calculations complemented by MLFF molecular dynamics. The optimized structures are dynamically stable within the harmonic approximation. Composition produces a clear electronic trend: MoS, MoSe, WS, and WSe are metallic, whereas MoTe and WTe are semiconducting, with HSE06 gaps of approximately 0.52 and 0.20 eV, respectively. The telluride gaps are additionally modulated by axial strain, demonstrating direct electromechanical tunability.

The effective axial stiffness decreases from sulfides to selenides and tellurides, consistent with increasing metal--chalcogen bond lengths. The reported Young's moduli, which depend on the adopted effective cross-sectional area, range from 392 to 718 GPa and should be interpreted as measures of axial stiffness rather than bending flexibility. All six nanowires have negative formation energies relative to the selected elemental reference phases; this result establishes energetic favorability with respect to those references but does not constitute a complete convex-hull stability analysis.

MLFF molecular-dynamics trajectories of 2.5 ns at 300 K preserve the one-dimensional morphology and well-defined local atomic order, providing evidence of finite-temperature dynamical stability over the simulated time scale. The curvature calculations reveal a crossover between approximately harmonic bending and a nonlinear regime characterized by nonuniform relaxation, including rippling and local buckling. The strong length dependence of the apparent bending rigidity, particularly for MoS, shows that the presently sampled geometries do not define a unique length-independent continuum bending constant. Instead, the calculations identify the onset of additional deformation modes as the wire length and imposed curvature increase. Direct DFT benchmarks of representative curved configurations are required before quantitative bending rigidities are assigned as intrinsic material constants. Overall, the results identify composition, axial strain, and curvature as complementary control parameters for the electronic and mechanical response of these ultrathin nanowires and motivate their further investigation in flexible nanoelectronic architectures.

\section*{Supporting Information}
Complete structural/phonon panels, orbital-resolved DOS/PDOS, charge-density maps, large-strain structural evolution, PBE band structures under axial strain, complete energy-- and stress--strain datasets, and MLFF energy traces are provided in the Supporting Information.

\section*{Acknowledgments}

The authors acknowledge the financial support  provided by the Brazilian funding agency CNPq under grant numbers 305174/2023-1, 408144/2022-0 and
305952/2023-4, and by CAPES through the PDPG-FAP Program under Project No. 88887.964834/2024-00, with scholarship support under Process No. 88887.184713/2025-00.  
Computational resources were provided by the Santos Dumont supercomputer at the National Laboratory for Scientific Computing (LNCC/MCTI, Brazil) and by the Laboratory of Advanced Scientific Computing (LaMCAD) at the Federal University of Goiás (UFG).

\bibliography{references}


{\Large Supporting Information for\\[0.5em]
Transition-Metal Chalcogenide Nanowires: A First-Principles Study of Stability, Electronic Structure, and Mechanical Response}

\setcounter{figure}{0}
\renewcommand{\thefigure}{S\arabic{figure}}
\renewcommand{\thesection}{S\arabic{section}}
\renewcommand{\thesubsection}{\thesection.\arabic{subsection}}

\section{Phonon dispersions}

The optimized geometries and projected equilibrium electronic band structures are presented in the main manuscript. Here we provide the phonon dispersions used to assess the dynamical stability of the six nanowires.

\begin{figure}[H]
\centering
\includegraphics[width=\linewidth,clip]{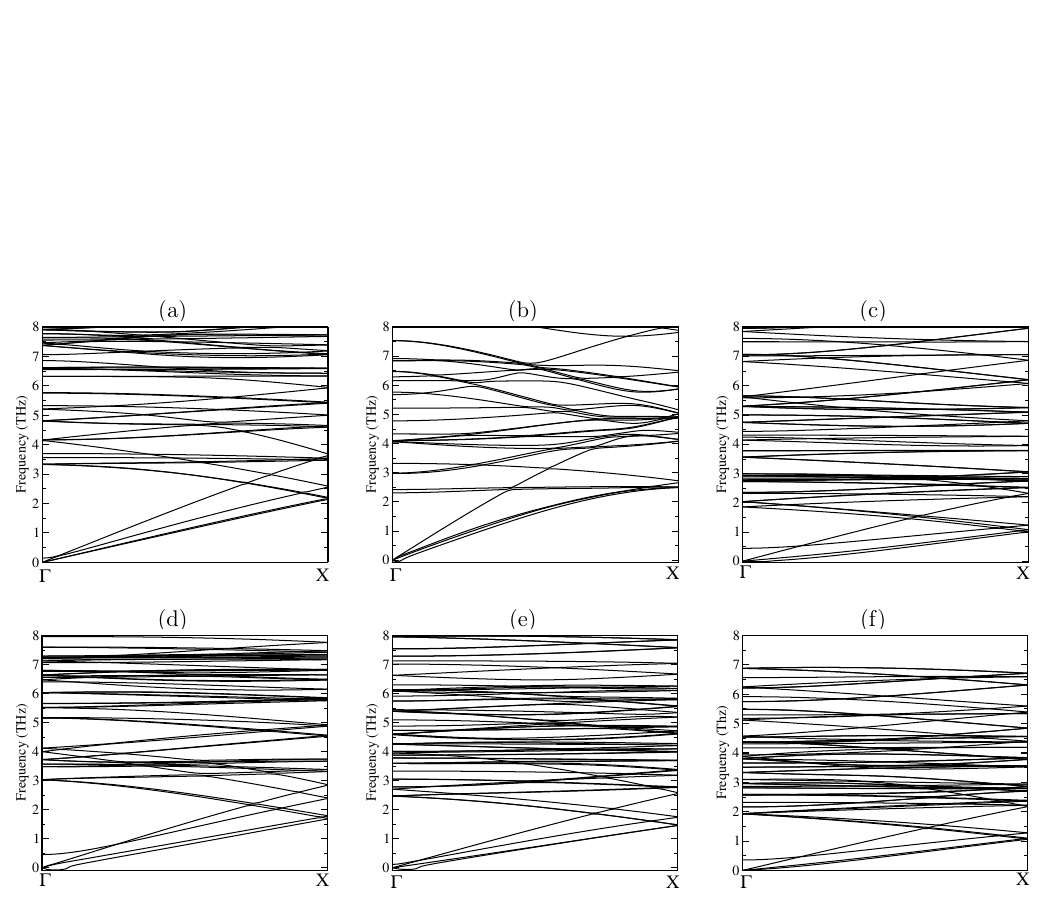}
\caption{Phonon dispersions of (a) MoS, (b) MoSe, (c) MoTe, (d) WS, (e) WSe, and (f) WTe nanowires. The absence of significant imaginary branches in the calculated dispersions supports the dynamical stability of the optimized structures within the numerical accuracy of the phonon calculations.}
\label{fig:si_phonons}
\end{figure}

\clearpage

\section{Orbital-resolved density of states}

\newcommand{\dospanel}[1]{%
\includegraphics[width=\linewidth,height=0.215\textheight,keepaspectratio,clip]{#1}%
}
\begin{figure}[H]
\centering
\includegraphics[width=\linewidth,clip]{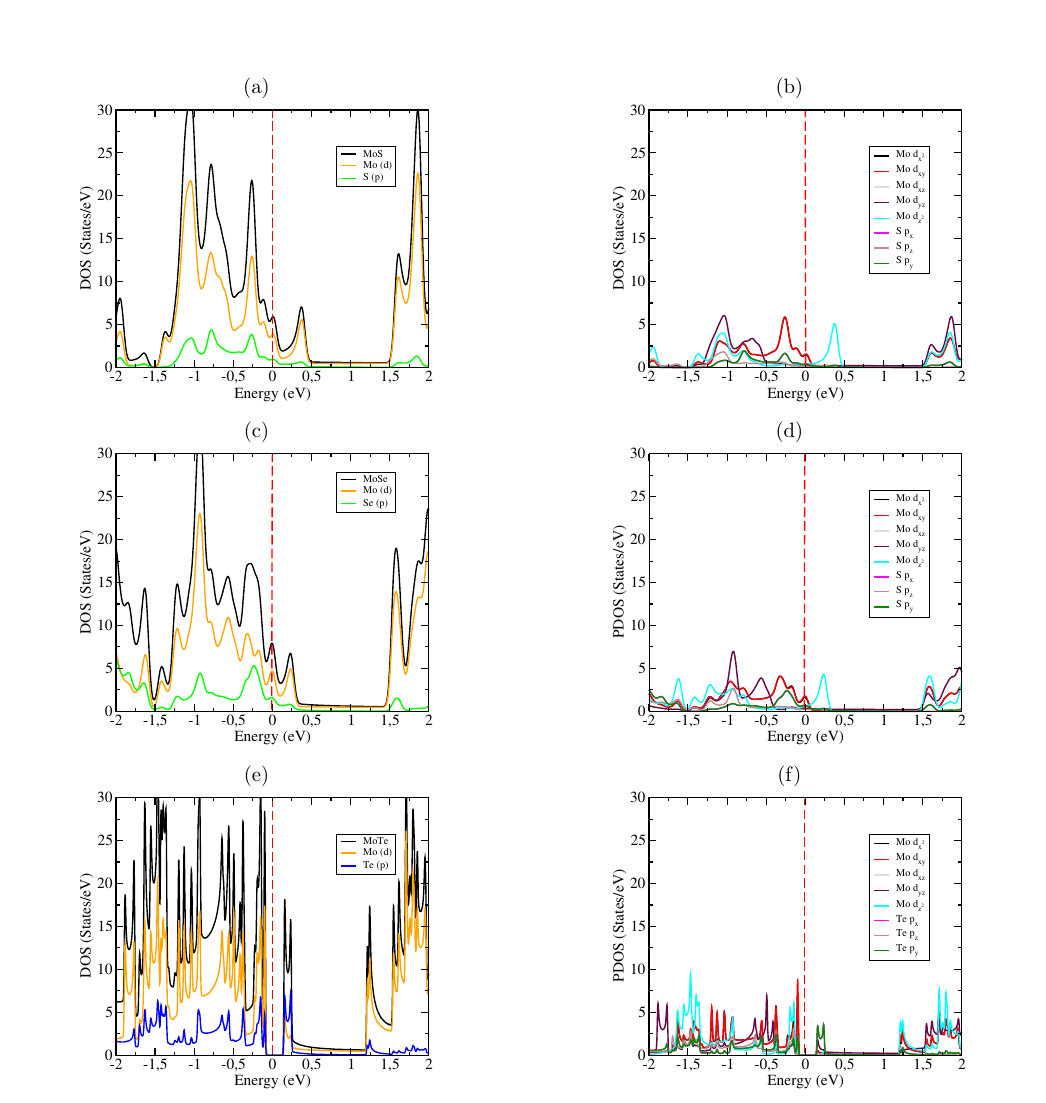}
\caption{Total and orbital-resolved density of states for the Mo-based nanowires.
For each material, the left panel shows the total DOS together with the projected
contributions from the transition-metal $d$ and chalcogen $p$ states, while the
right panel shows the orbital decomposition of the Mo $d$ and chalcogen $p$
states. From top to bottom: MoS, MoSe, and MoTe. The Fermi level is set to zero.}
\label{si:fig_dos_mo}
\end{figure}


\begin{figure}[H]
\centering
\includegraphics[width=\linewidth,clip]{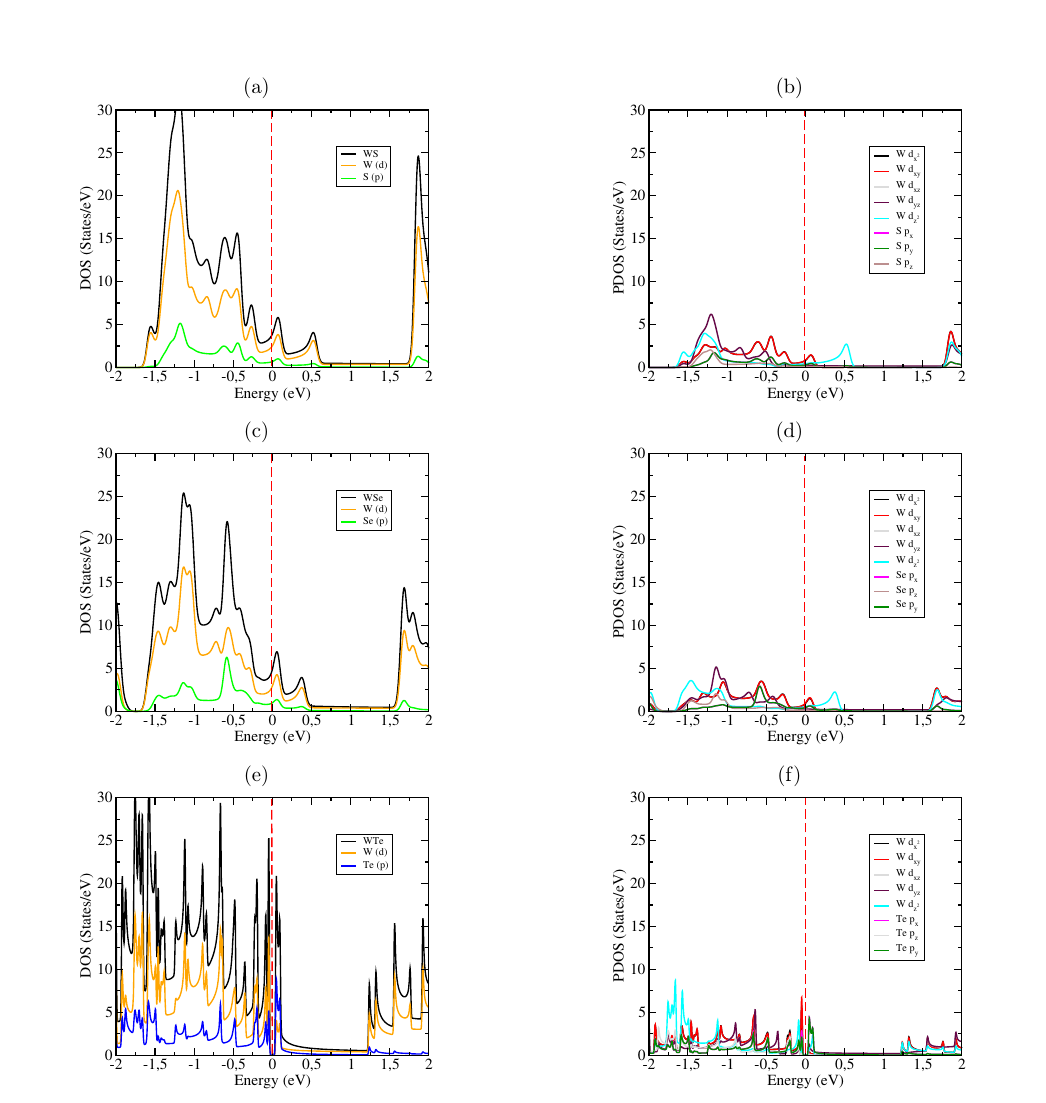}
\caption{Total and orbital-resolved density of states for the W-based nanowires.
For each material, the left panel shows the total DOS together with the projected
contributions from the transition-metal $d$ and chalcogen $p$ states, while the
right panel shows the orbital decomposition of the W $d$ and chalcogen $p$
states. From top to bottom: WS, WSe, and WTe. The Fermi level is set to zero.}
\label{si:fig_dos_w}
\end{figure}

\clearpage

\section{Large-strain structural evolution}

\begin{figure}[tb]
\centering
\includegraphics[width=\linewidth,clip]{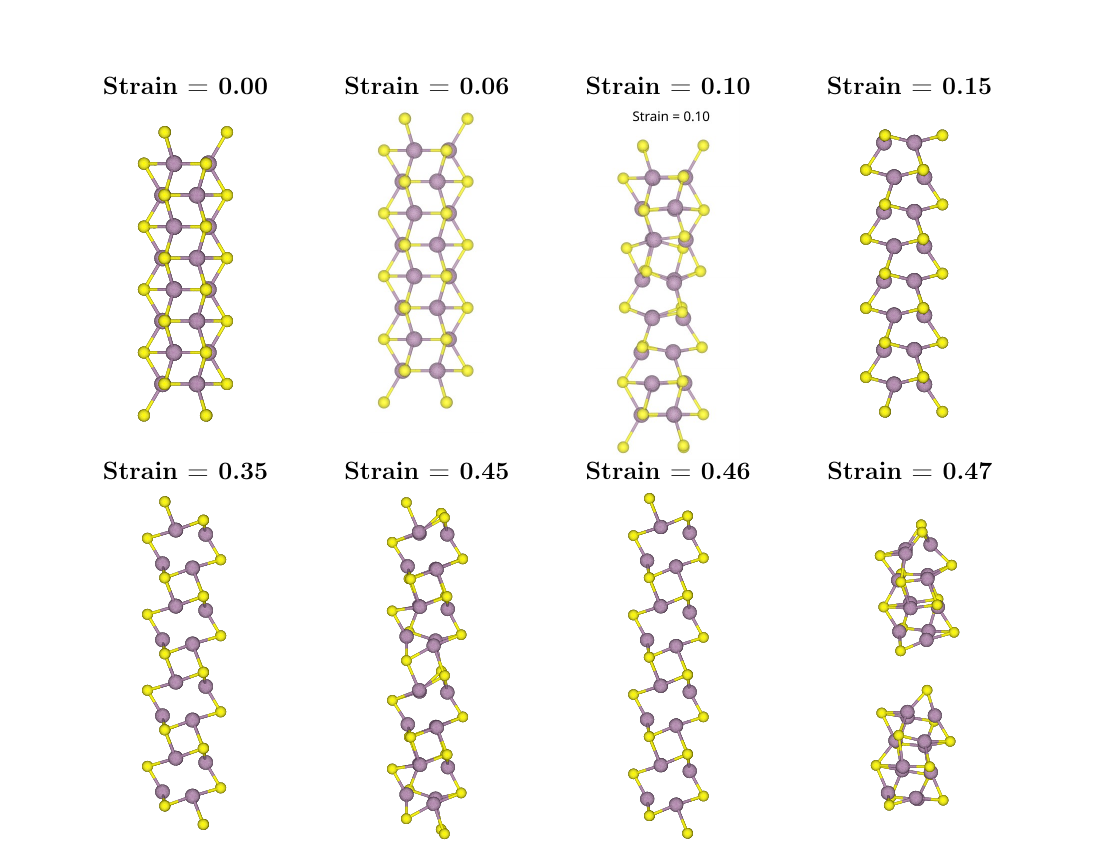}

\caption{Different stages during the stretching process of a pristine MoS nanowire until rupture.}
\label{si:fig_MoS_strain}
\end{figure}

\begin{figure}[H]
\centering
\includegraphics[width=\linewidth,clip]{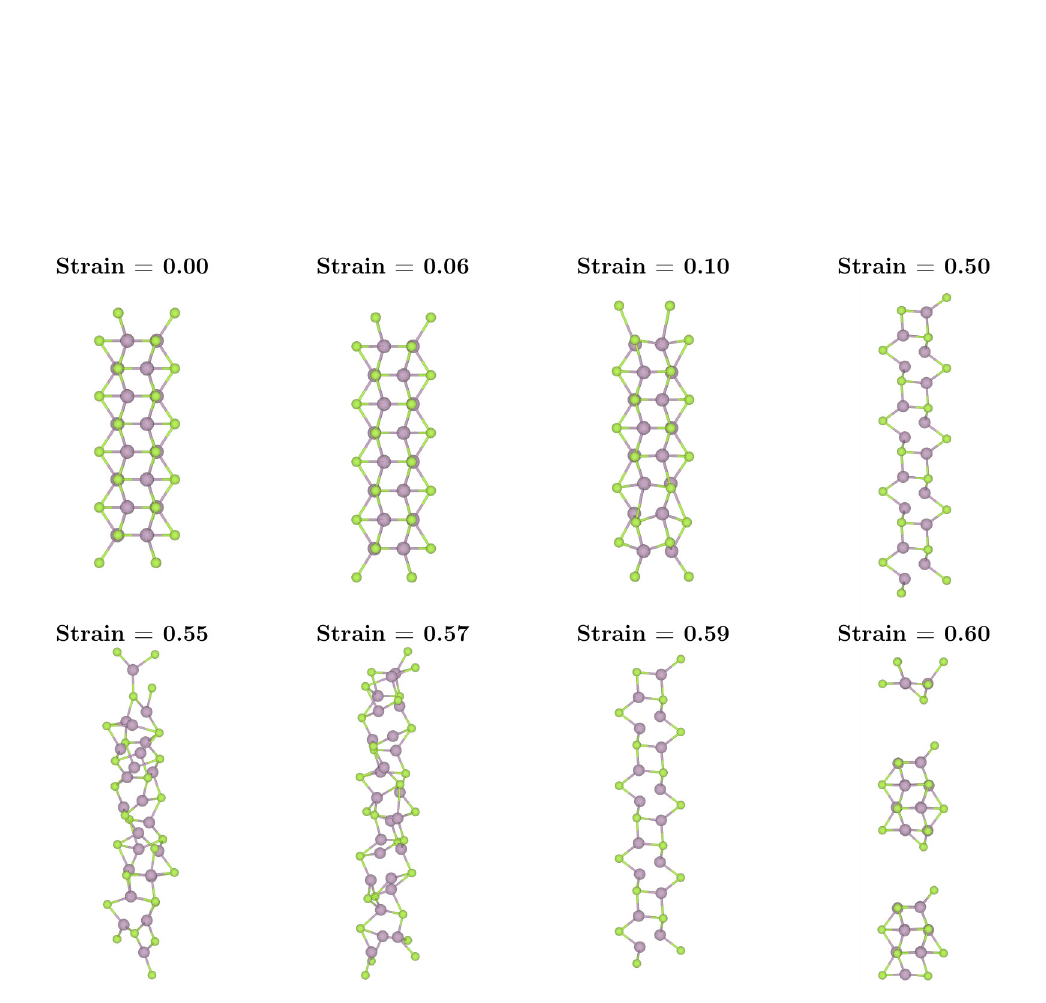}
\caption{Different stages during the stretching process of a pristine MoSe nanowire until rupture.}
\label{si:fig_MoSe_strain}
\end{figure}

\begin{figure}[H]
\centering
\includegraphics[width=\linewidth,clip]{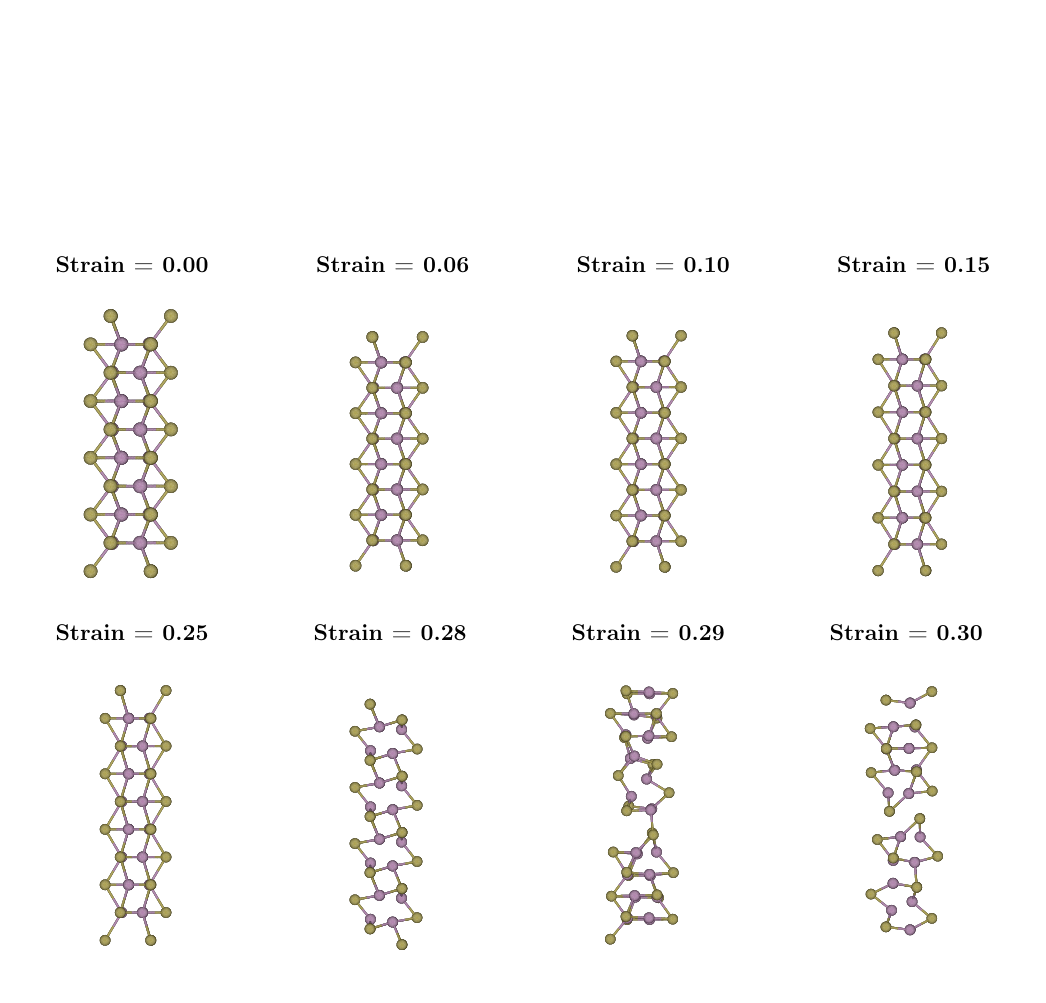}
\caption{Different stages during the stretching process of a pristine MoTe nanowire until rupture.}
\label{si:fig_MoTe_strain}
\end{figure}

\begin{figure}[H]
\centering
\includegraphics[width=\linewidth,clip]{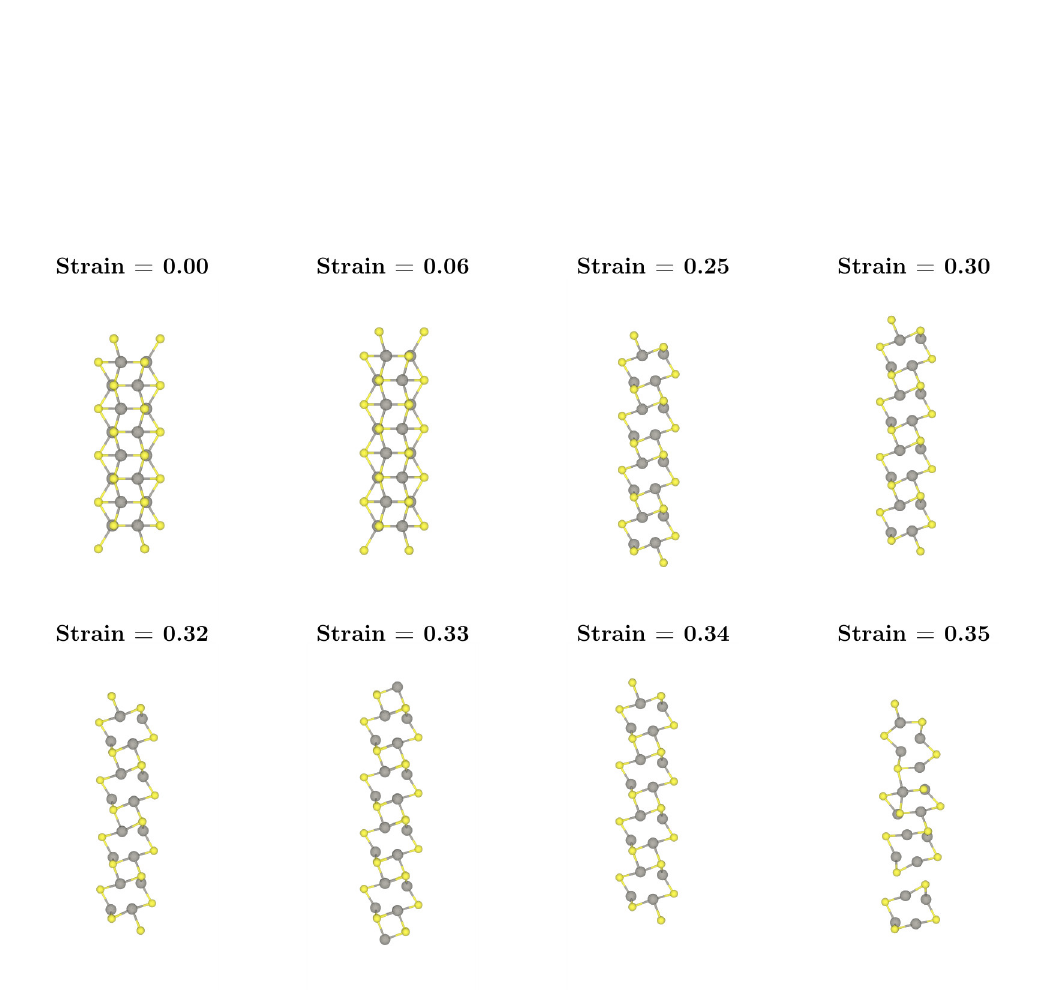}
\caption{Different stages during the stretching process of a pristine WS nanowire until rupture.}
\label{si:fig_WS_strain}
\end{figure}

\begin{figure}[H]
\centering
\includegraphics[width=\linewidth,clip]{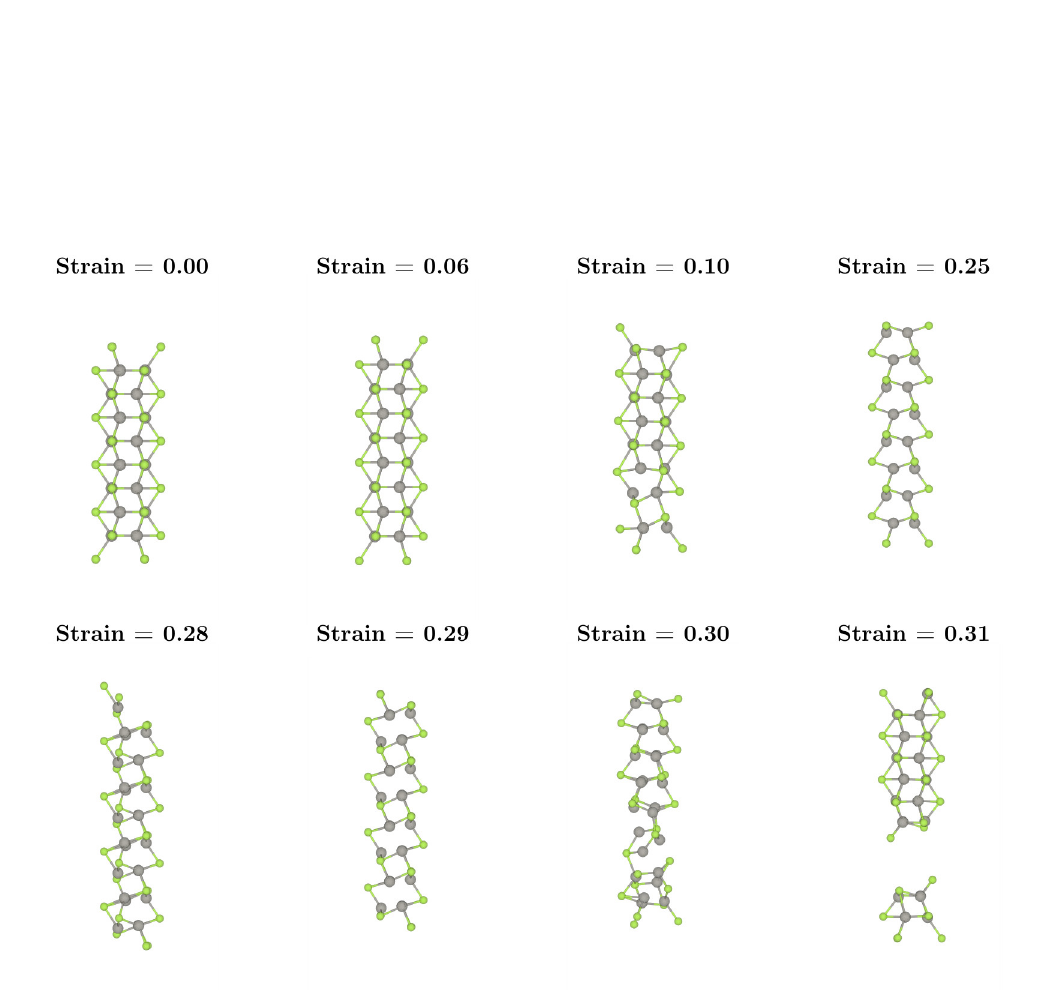}
\caption{Different stages during the stretching process of a pristine WSe nanowire until rupture.}
\label{si:fig_WSe_strain}
\end{figure}

\begin{figure}[tb]
\centering
\includegraphics[width=\linewidth,clip]{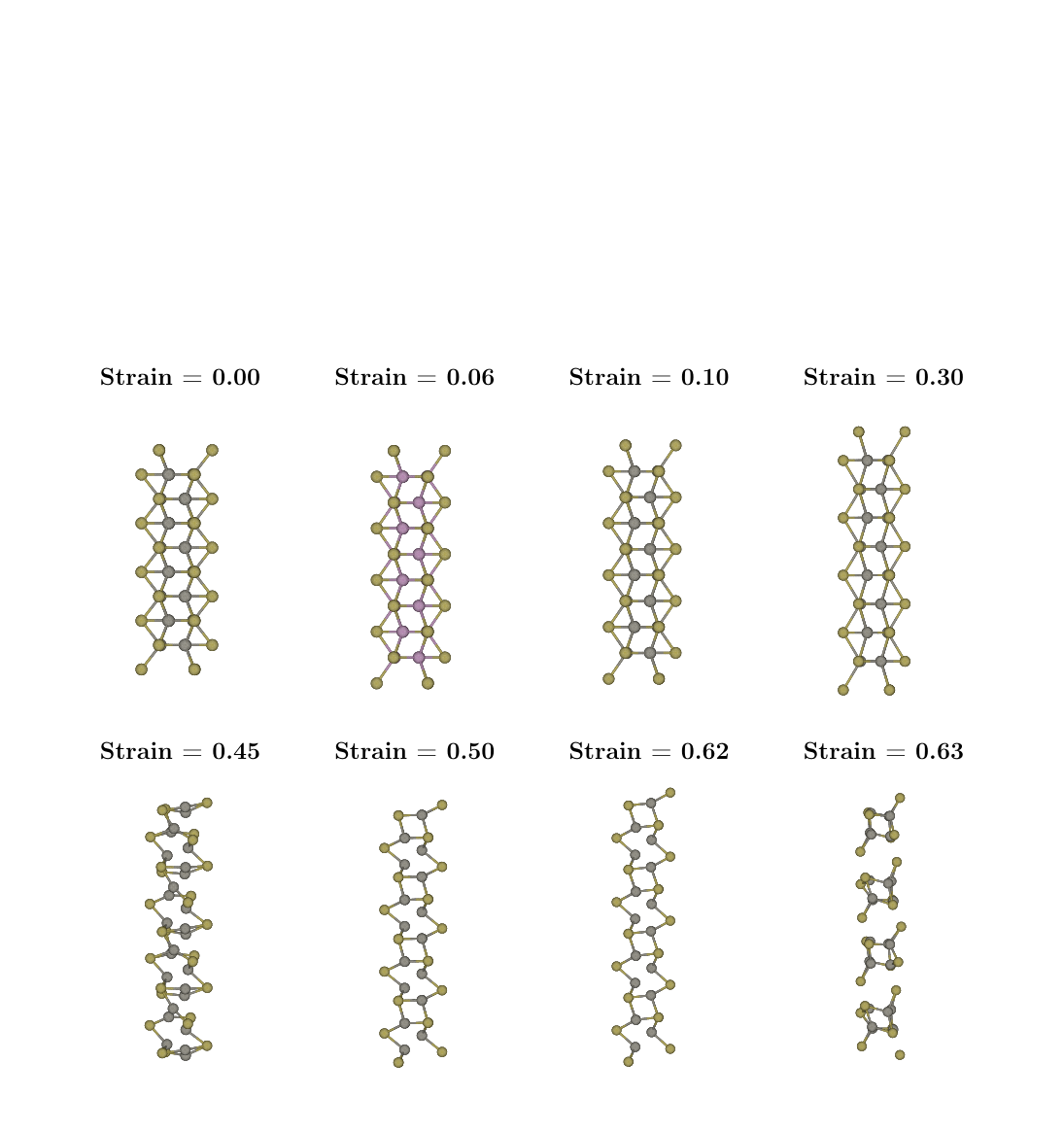}
\caption{Different stages during the stretching process of a pristine WTe nanowire until rupture.}
\label{si:fig_WTe_strain}
\end{figure}

\clearpage

\section{Electronic band structures under axial strain}

\begin{figure}[H]
\centering
\includegraphics[width=\linewidth,clip]{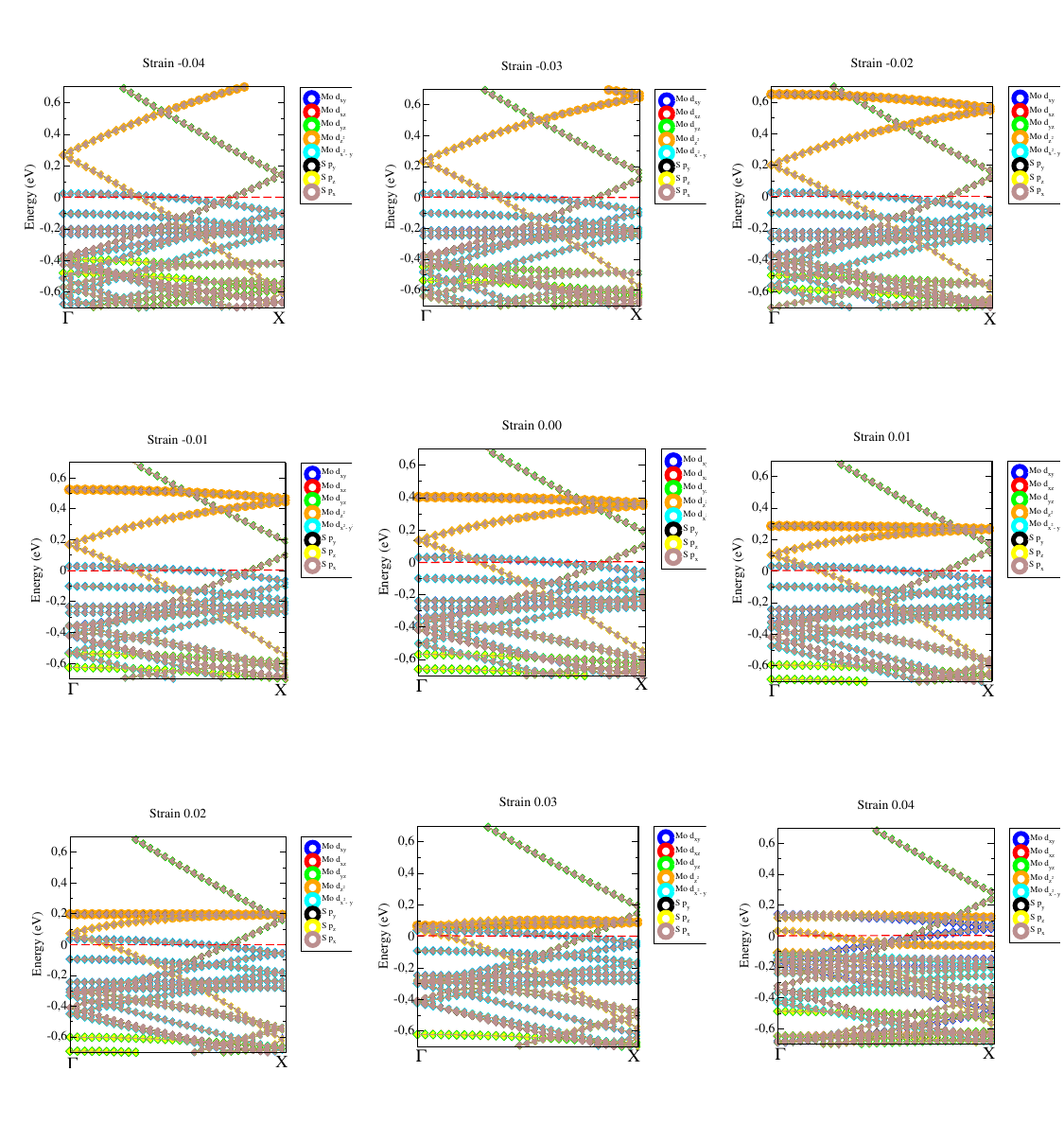}
\caption{Orbital-projected electronic band structures calculated within DFT-PBE including spin--orbit coupling (SOC) for the pristine MoS nanowire for the set of axial strains shown, spanning approximately $-0.04$ to $0.04$, using a $1\times1\times4$ supercell. The dotted lines indicate the Fermi level.}
\label{si:bands_MoS}
\end{figure}

\begin{figure}[H]
\centering
\includegraphics[width=\linewidth,clip]{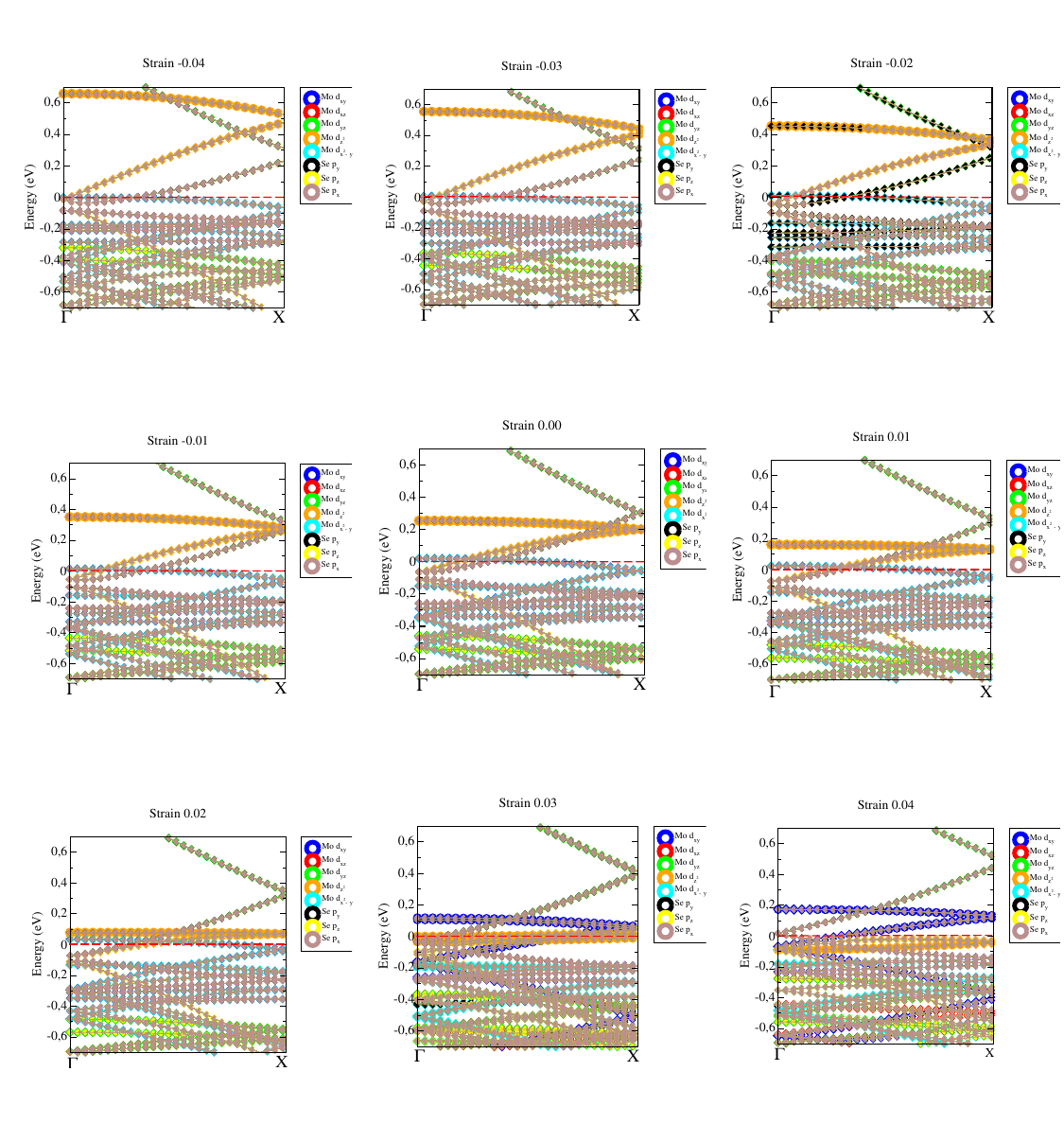}
\caption{Orbital-projected electronic band structures calculated within DFT-PBE including spin--orbit coupling (SOC) for the pristine MoSe nanowire for the set of axial strains shown, spanning approximately $-0.04$ to $0.04$, using a $1\times1\times4$ supercell. The dotted lines indicate the Fermi level.}
\label{si:bands_MoSe}
\end{figure}

\begin{figure}[H]
\centering
\includegraphics[width=\linewidth,clip]{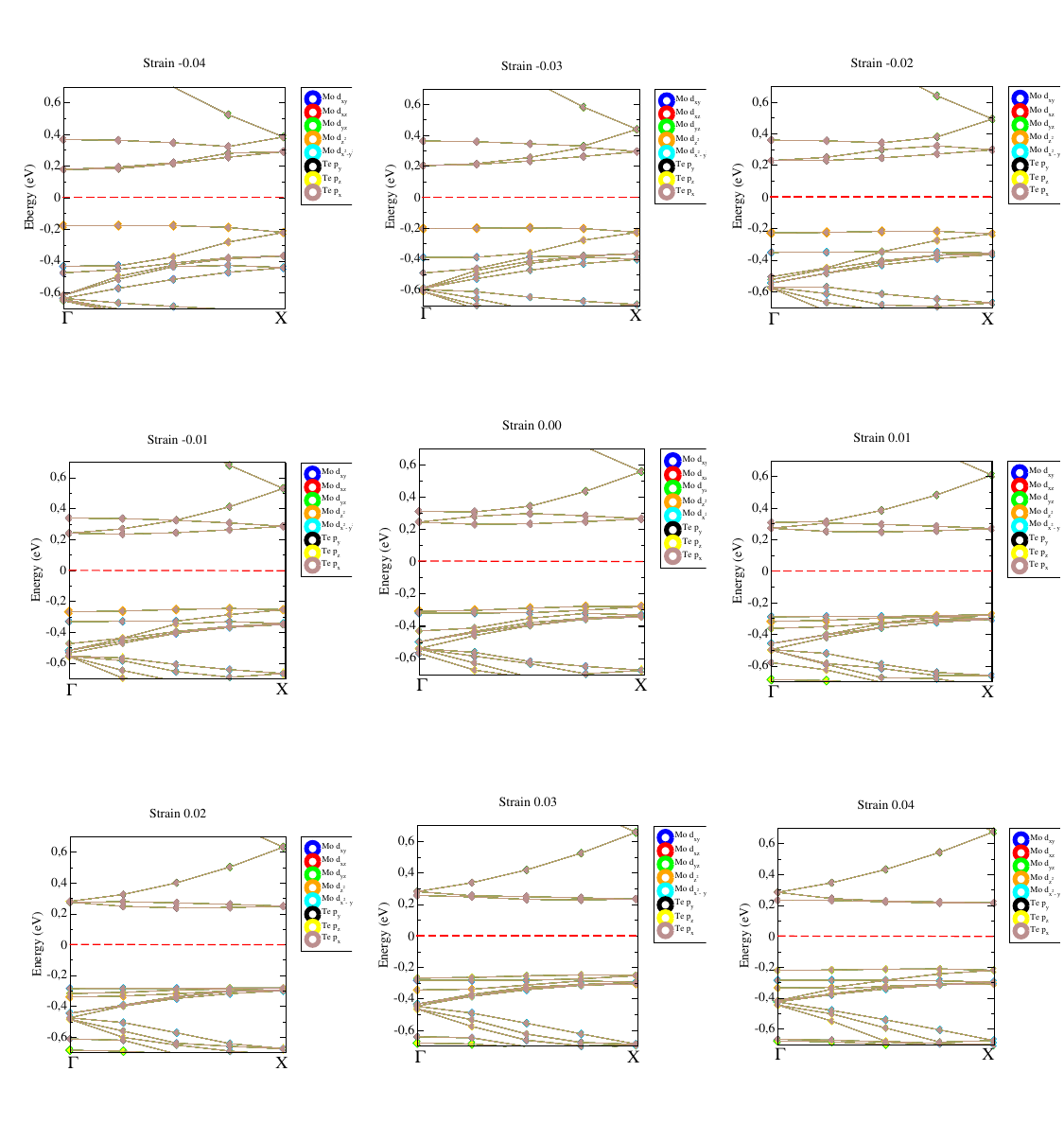}
\caption{Orbital-projected electronic band structures calculated within DFT-PBE including spin--orbit coupling (SOC) for the pristine MoTe nanowire for the set of axial strains shown, spanning approximately $-0.04$ to $0.04$, using a $1\times1\times4$ supercell. The dotted lines indicate the Fermi level.}
\label{si:bands_MoTe}
\end{figure}

\begin{figure}[H]
\centering
\includegraphics[width=\linewidth,clip]{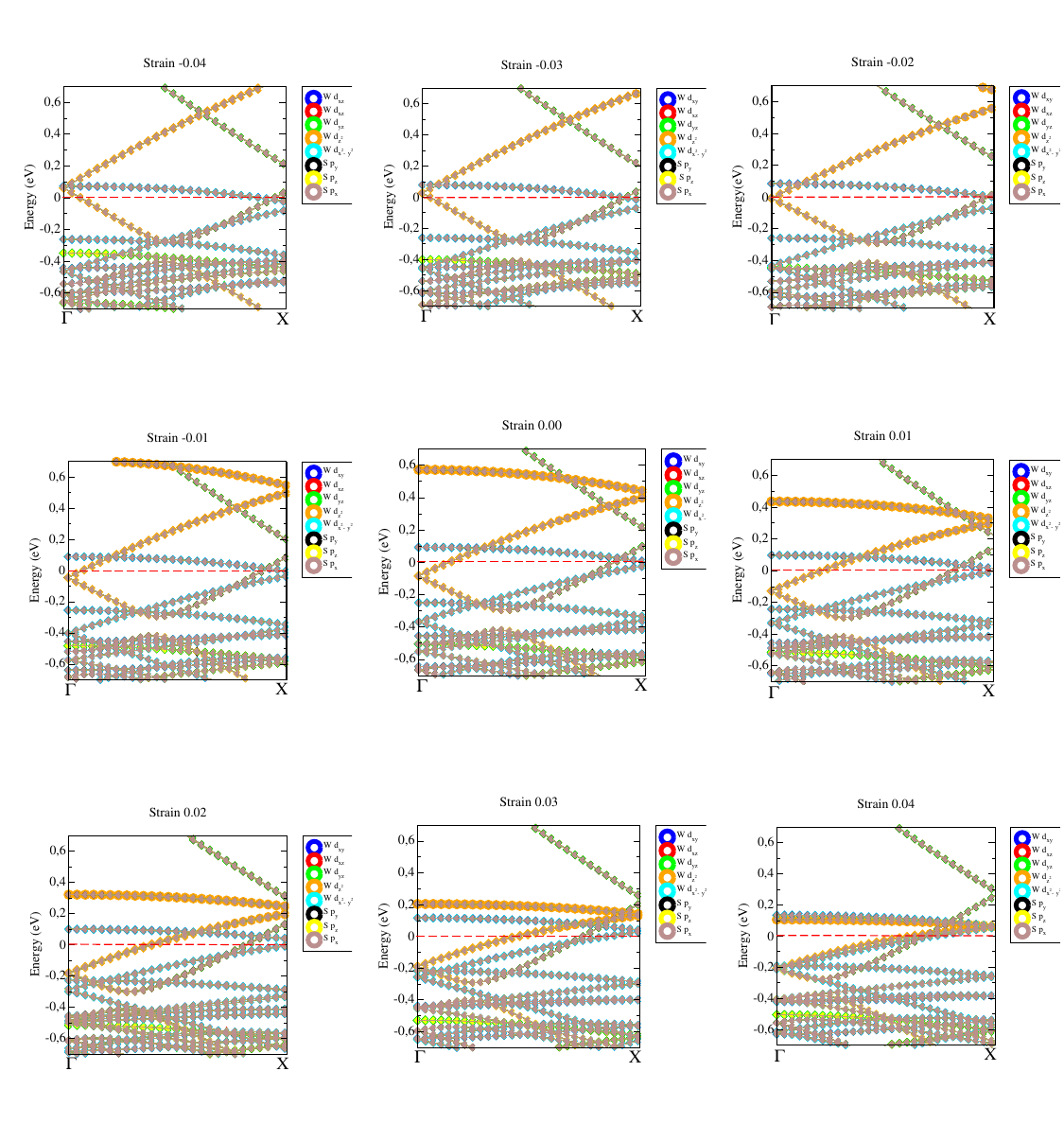}
\caption{Orbital-projected electronic band structures calculated within DFT-PBE including spin--orbit coupling (SOC) for the pristine WS nanowire for the set of axial strains shown, spanning approximately $-0.04$ to $0.04$, using a $1\times1\times4$ supercell. The dotted lines indicate the Fermi level.}
\label{si:bands_WS}
\end{figure}

\begin{figure}[H]
\centering
\includegraphics[width=\linewidth,clip]{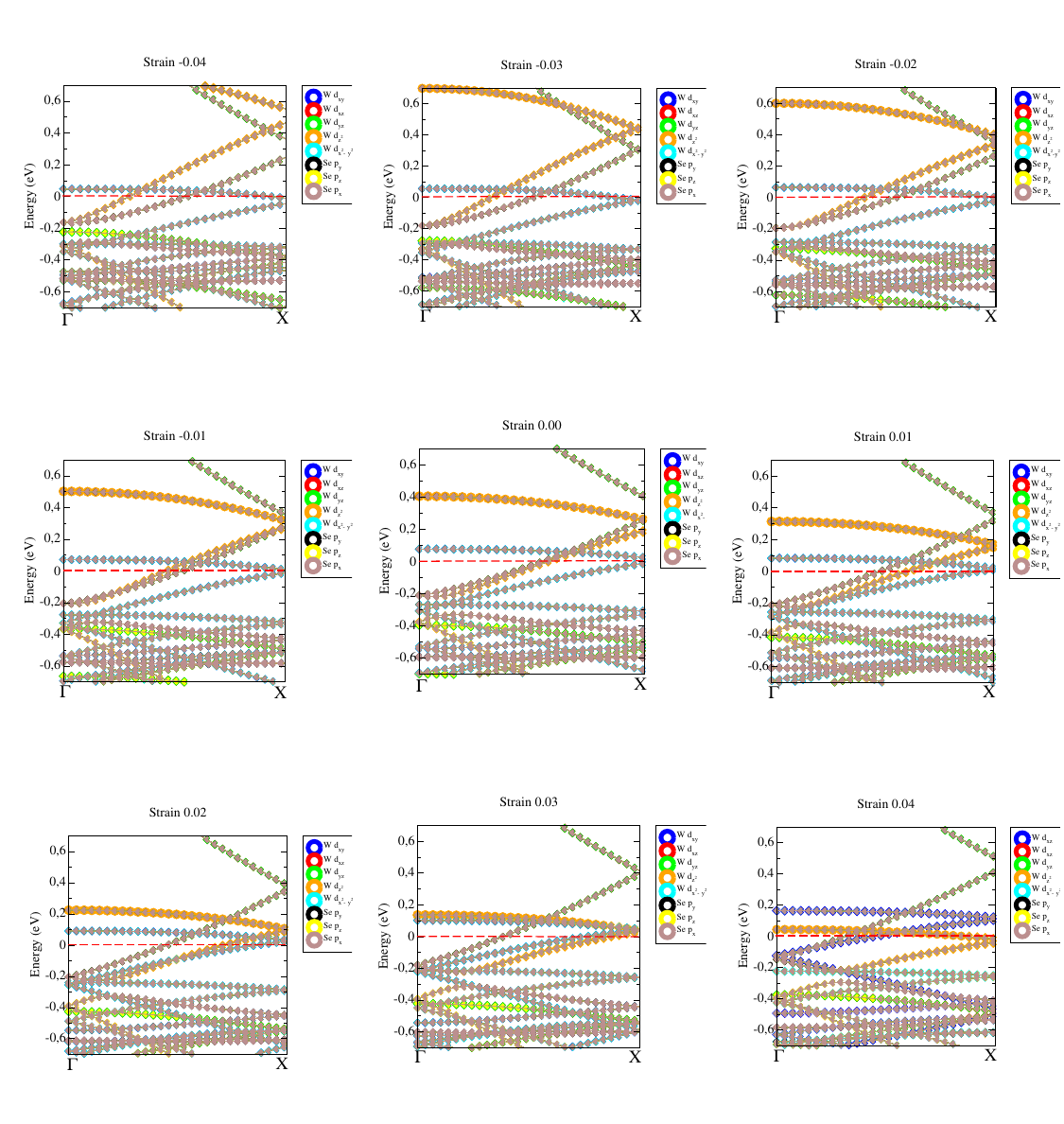}
\caption{Orbital-projected electronic band structures calculated within DFT-PBE including spin--orbit coupling (SOC) for the pristine WSe nanowire for the set of axial strains shown, spanning approximately $-0.04$ to $0.04$, using a $1\times1\times4$ supercell. The dotted lines indicate the Fermi level.}
\label{si:bands_WSe}
\end{figure}

\begin{figure}[H]
\centering
\includegraphics[width=\linewidth,clip]{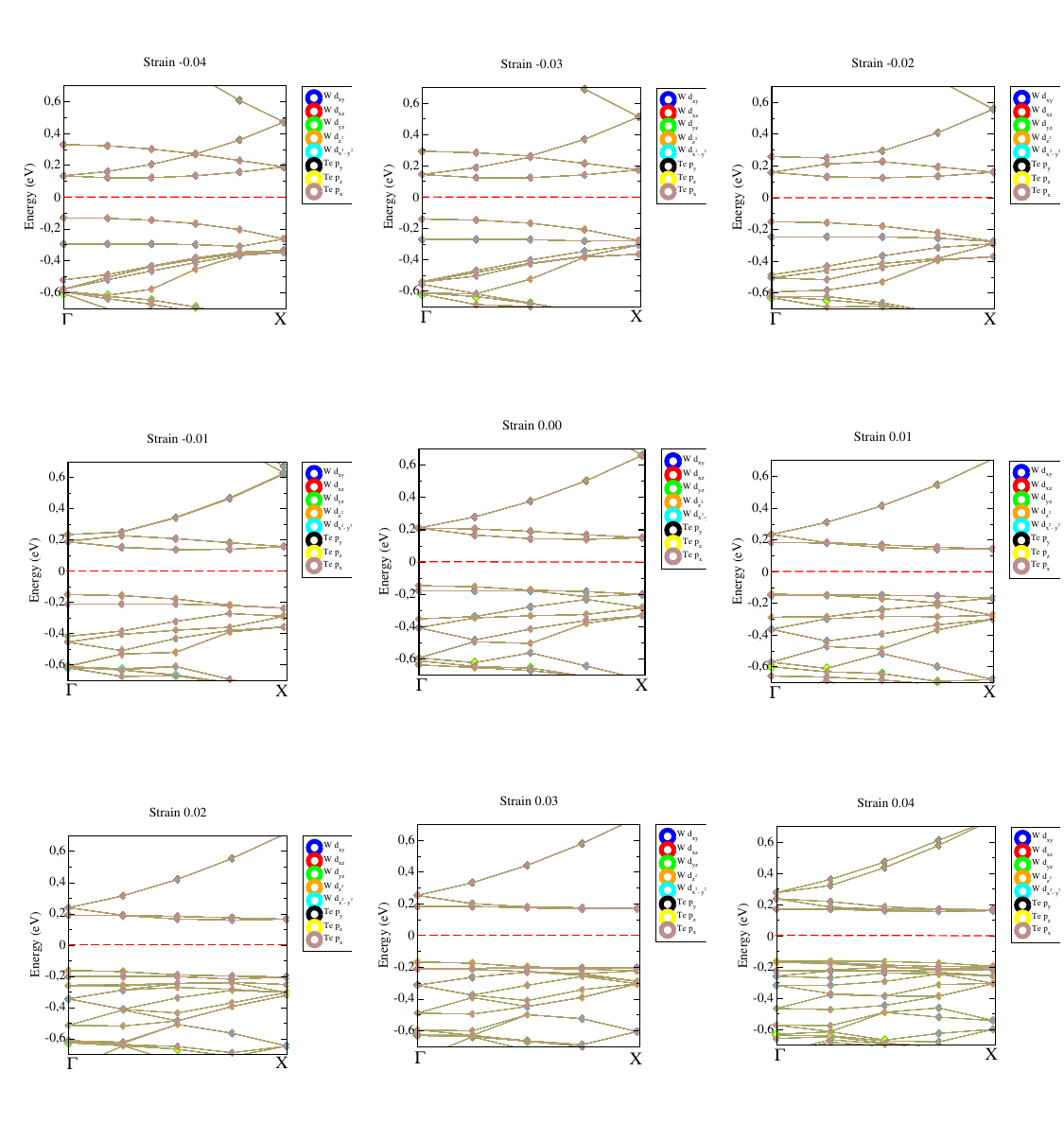}
\caption{Orbital-projected electronic band structures calculated within DFT-PBE including spin--orbit coupling (SOC) for the pristine WTe nanowire for the set of axial strains shown, spanning approximately $-0.04$ to $0.04$, using a $1\times1\times4$ supercell. The dotted lines indicate the Fermi level.}
\label{si:bands_WTe}
\end{figure}

\clearpage

\section{Complete energy-- and stress--strain curves}


\begin{figure}[H]
\centering
\includegraphics[width=\linewidth,clip]{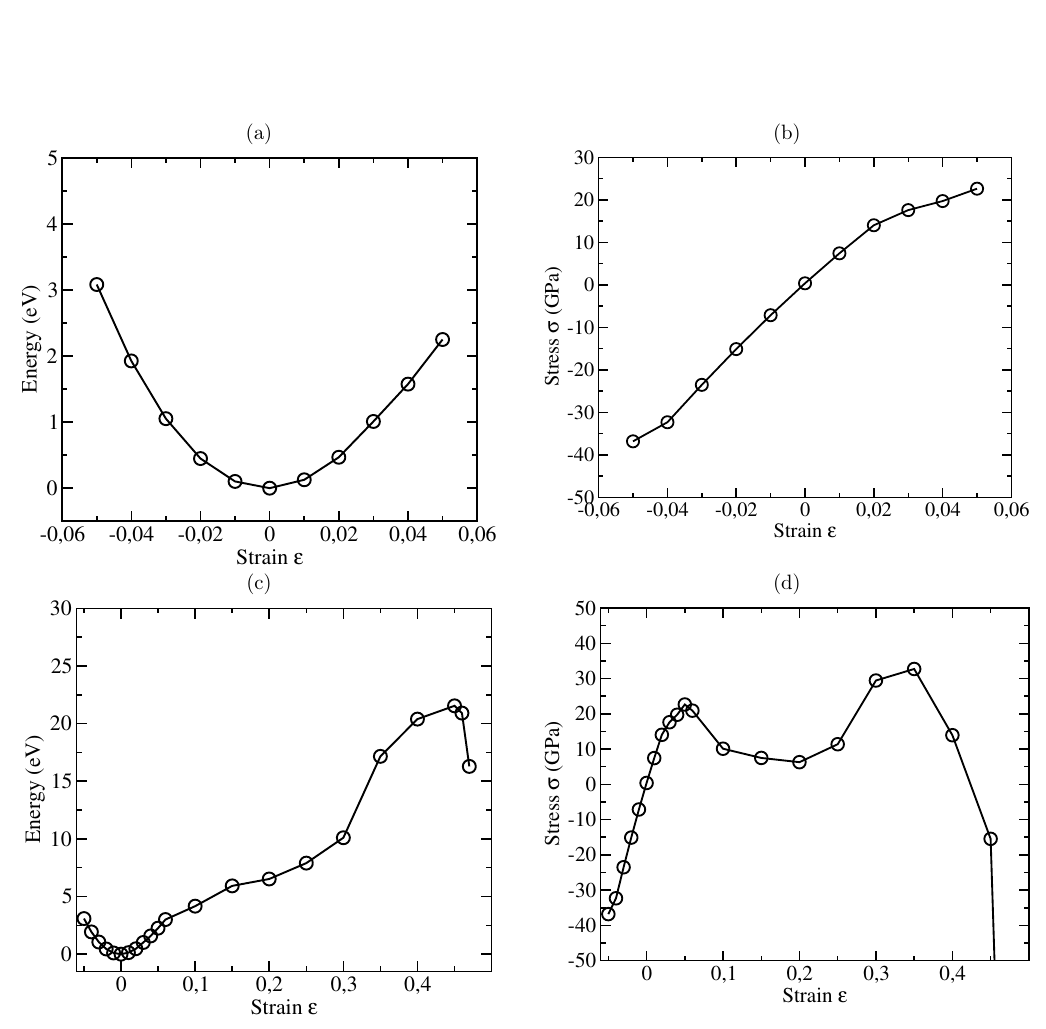}
\caption{Axial mechanical response of the pristine MoS nanowire. (a) Near-equilibrium total energy as a function of strain, referenced to the zero-strain configuration; (b) corresponding near-equilibrium stress--strain curve; (c) extended-range total-energy curve approaching large deformation and rupture; and (d) corresponding extended-range stress--strain curve. The small-strain data are used for the elastic analysis in the main manuscript.}
\label{si:mechanical_MoS}
\end{figure}

\begin{figure}[H]
\centering
\includegraphics[width=\linewidth,clip]{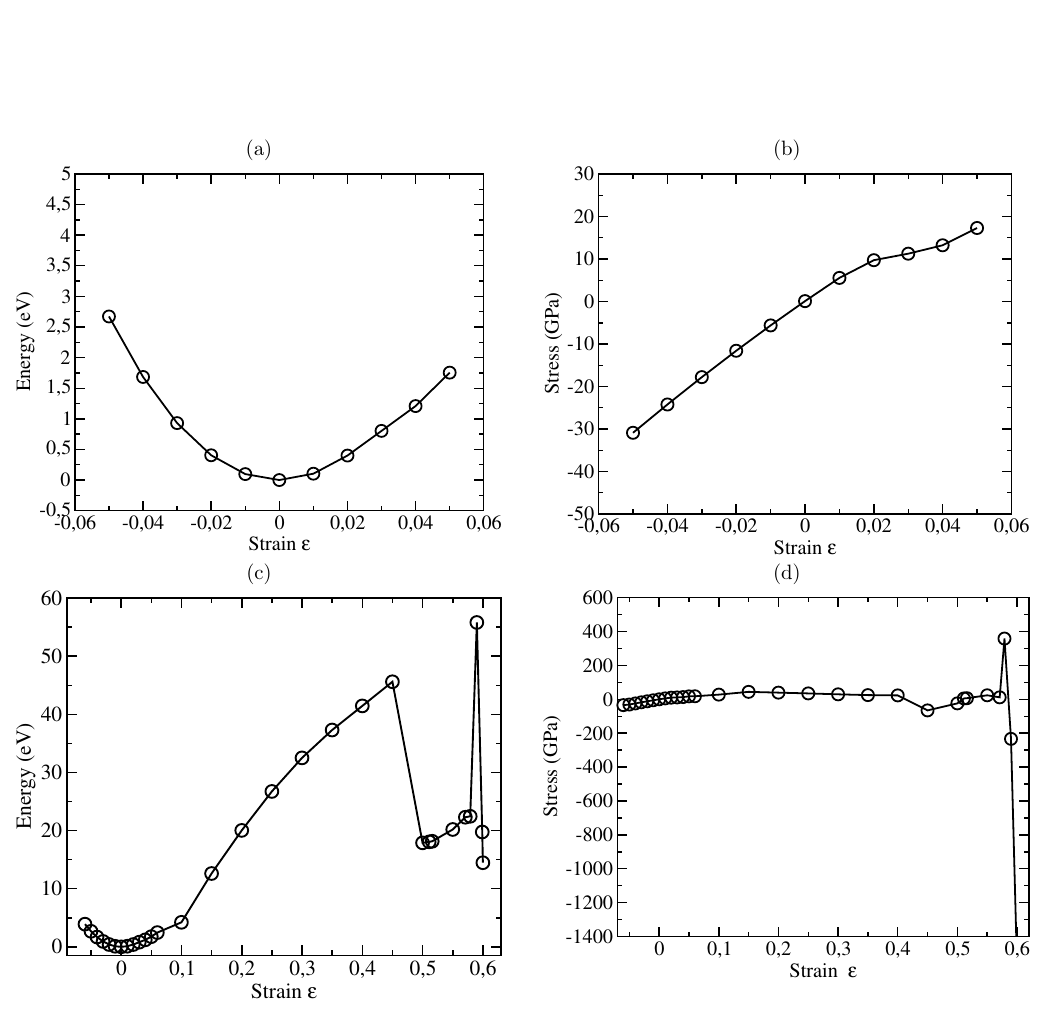}
\caption{Axial mechanical response of the pristine MoSe nanowire. (a) Near-equilibrium total energy as a function of strain, referenced to the zero-strain configuration; (b) corresponding near-equilibrium stress--strain curve; (c) extended-range total-energy curve approaching large deformation and rupture; and (d) corresponding extended-range stress--strain curve. The small-strain data are used for the elastic analysis in the main manuscript.}
\label{si:mechanical_MoSe}
\end{figure}

\begin{figure}[H]
\centering
\includegraphics[width=\linewidth,clip]{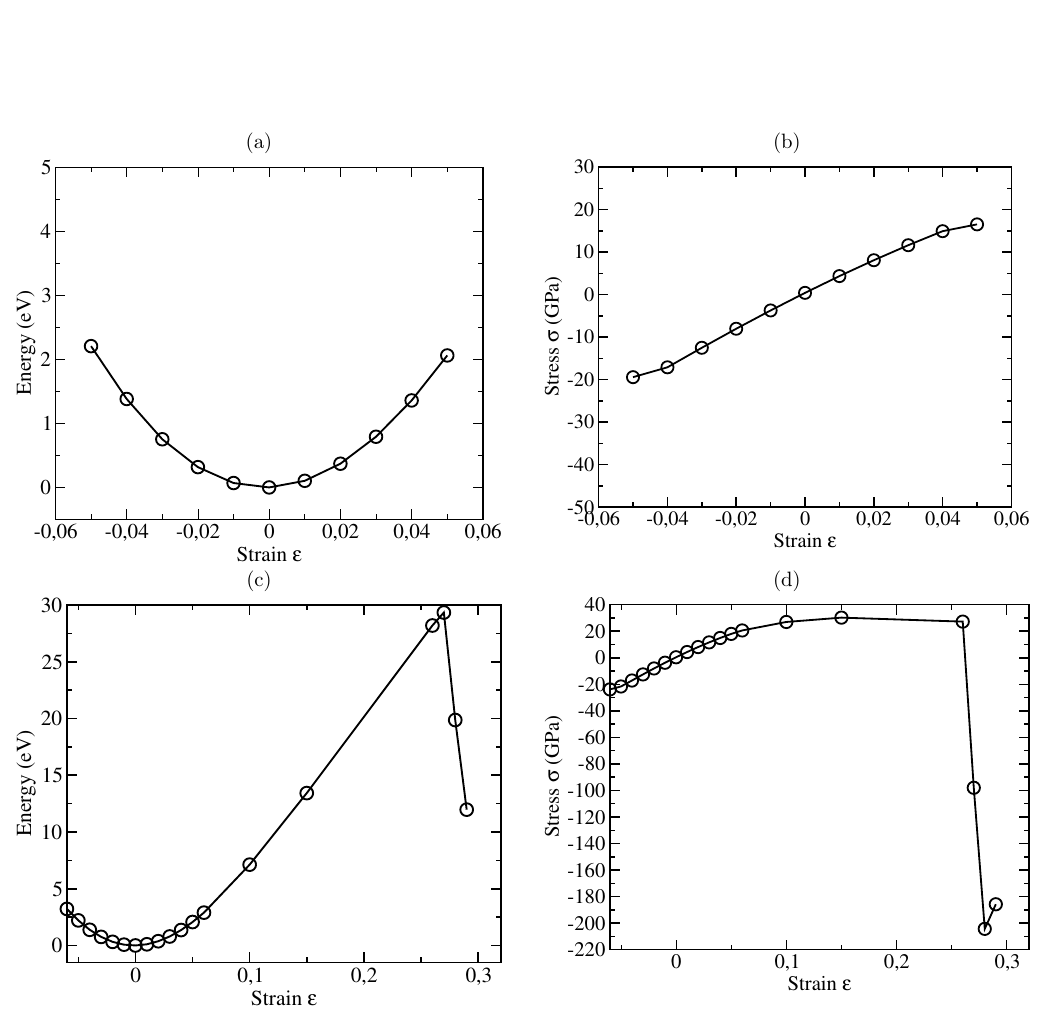}
\caption{Axial mechanical response of the pristine MoTe nanowire. (a) Near-equilibrium total energy as a function of strain, referenced to the zero-strain configuration; (b) corresponding near-equilibrium stress--strain curve; (c) extended-range total-energy curve approaching large deformation and rupture; and (d) corresponding extended-range stress--strain curve. The small-strain data are used for the elastic analysis in the main manuscript.}
\label{si:mechanical_MoTe}
\end{figure}

\begin{figure}[H]
\centering
\includegraphics[width=\linewidth,clip]{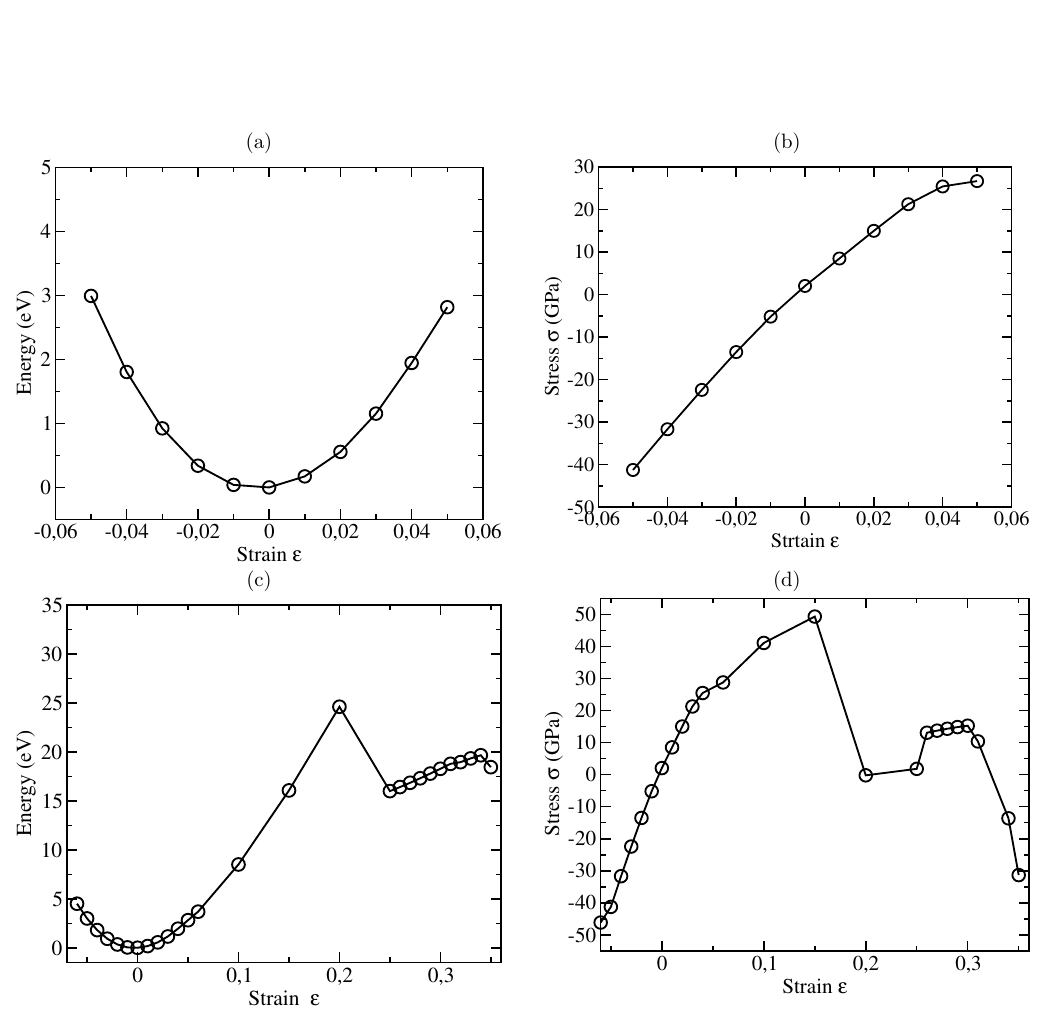}
\caption{Axial mechanical response of the pristine WS nanowire. (a) Near-equilibrium total energy as a function of strain, referenced to the zero-strain configuration; (b) corresponding near-equilibrium stress--strain curve; (c) extended-range total-energy curve approaching large deformation and rupture; and (d) corresponding extended-range stress--strain curve. The small-strain data are used for the elastic analysis in the main manuscript.}
\label{si:mechanical_WS}
\end{figure}

\begin{figure}[H]
\centering
\includegraphics[width=\linewidth,clip]{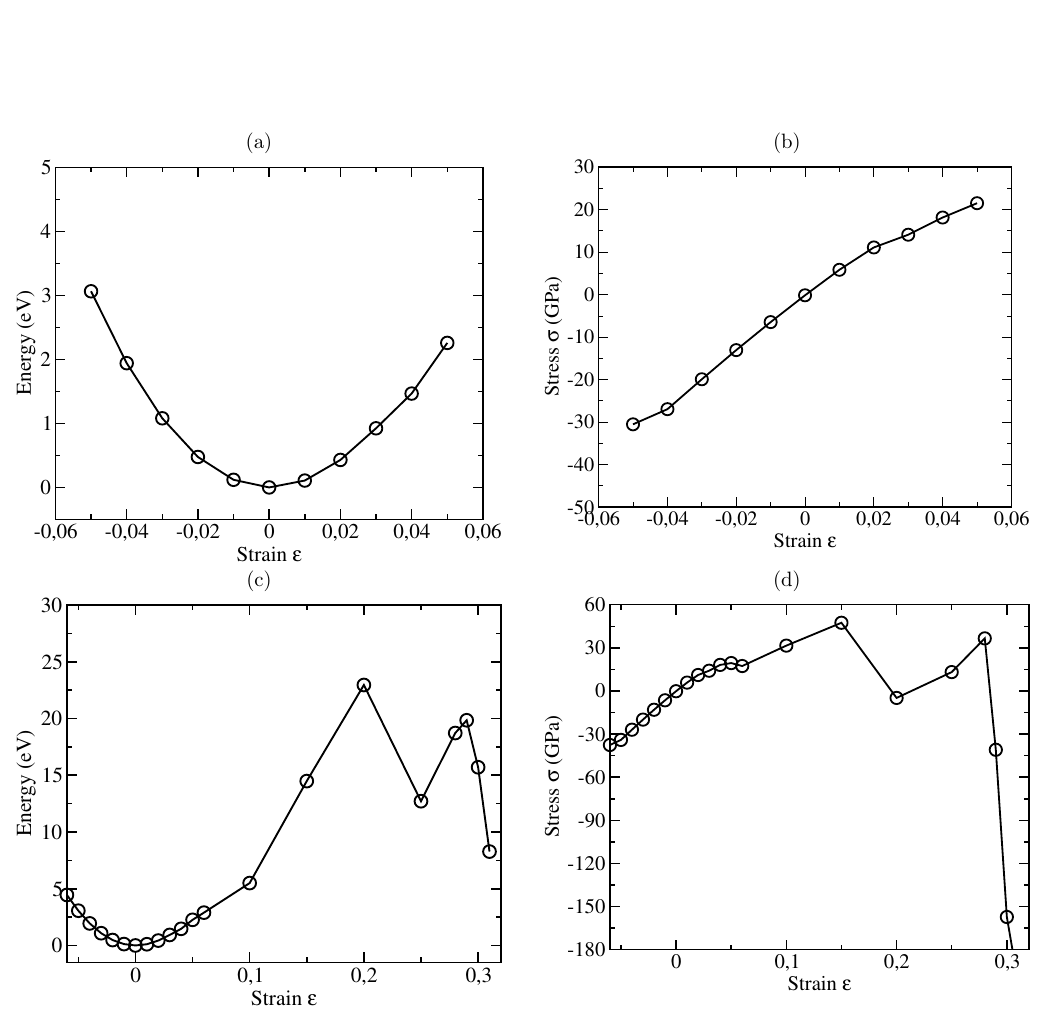}
\caption{Axial mechanical response of the pristine WSe nanowire. (a) Near-equilibrium total energy as a function of strain, referenced to the zero-strain configuration; (b) corresponding near-equilibrium stress--strain curve; (c) extended-range total-energy curve approaching large deformation and rupture; and (d) corresponding extended-range stress--strain curve. The small-strain data are used for the elastic analysis in the main manuscript.}
\label{si:mechanical_WSe}
\end{figure}

\begin{figure}[H]
\centering
\includegraphics[width=\linewidth,clip]{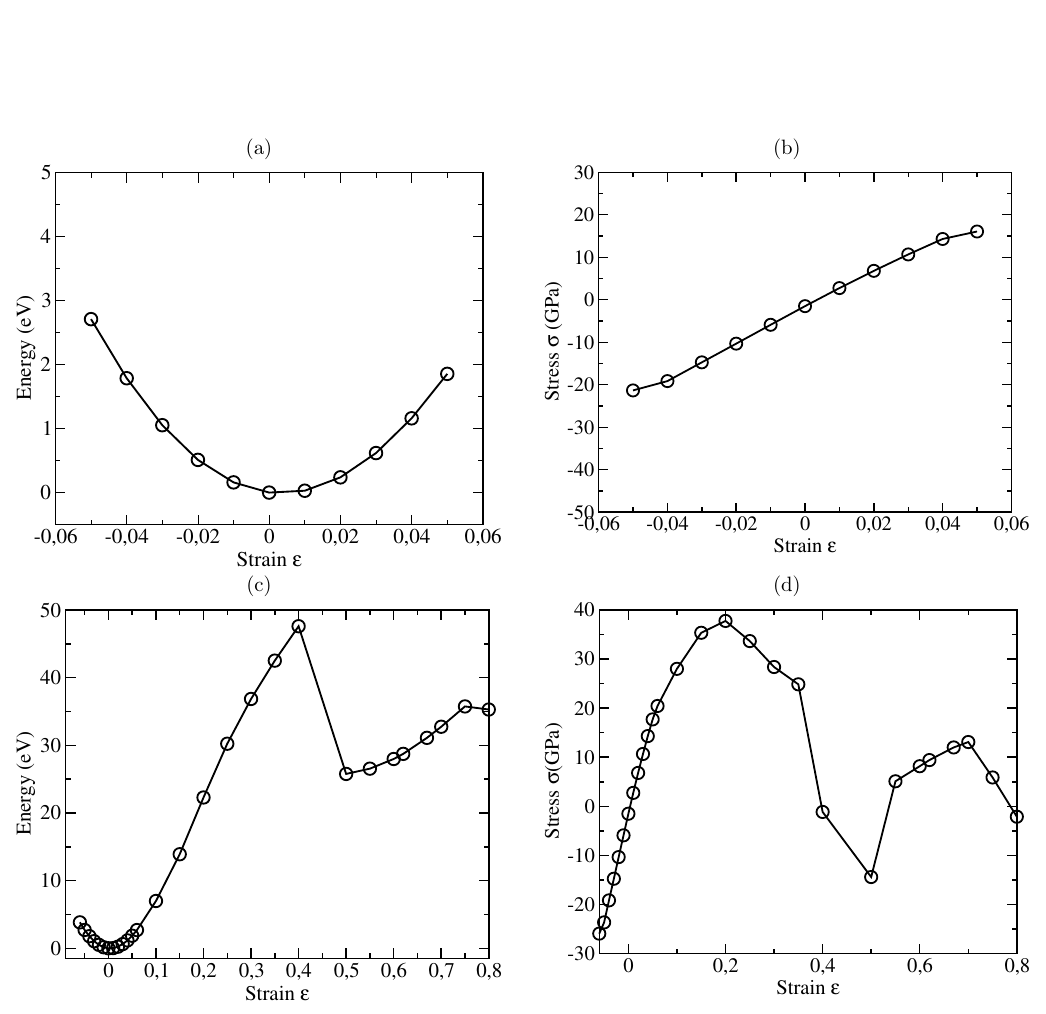}
\caption{Axial mechanical response of the pristine WTe nanowire. (a) Near-equilibrium total energy as a function of strain, referenced to the zero-strain configuration; (b) corresponding near-equilibrium stress--strain curve; (c) extended-range total-energy curve approaching large deformation and rupture; and (d) corresponding extended-range stress--strain curve. The small-strain data are used for the elastic analysis in the main manuscript.}
\label{si:mechanical_WTe}
\end{figure}

\clearpage

\section{MLFF molecular-dynamics energy traces}

\begin{figure}[H]
\centering
\includegraphics[width=\linewidth,clip]{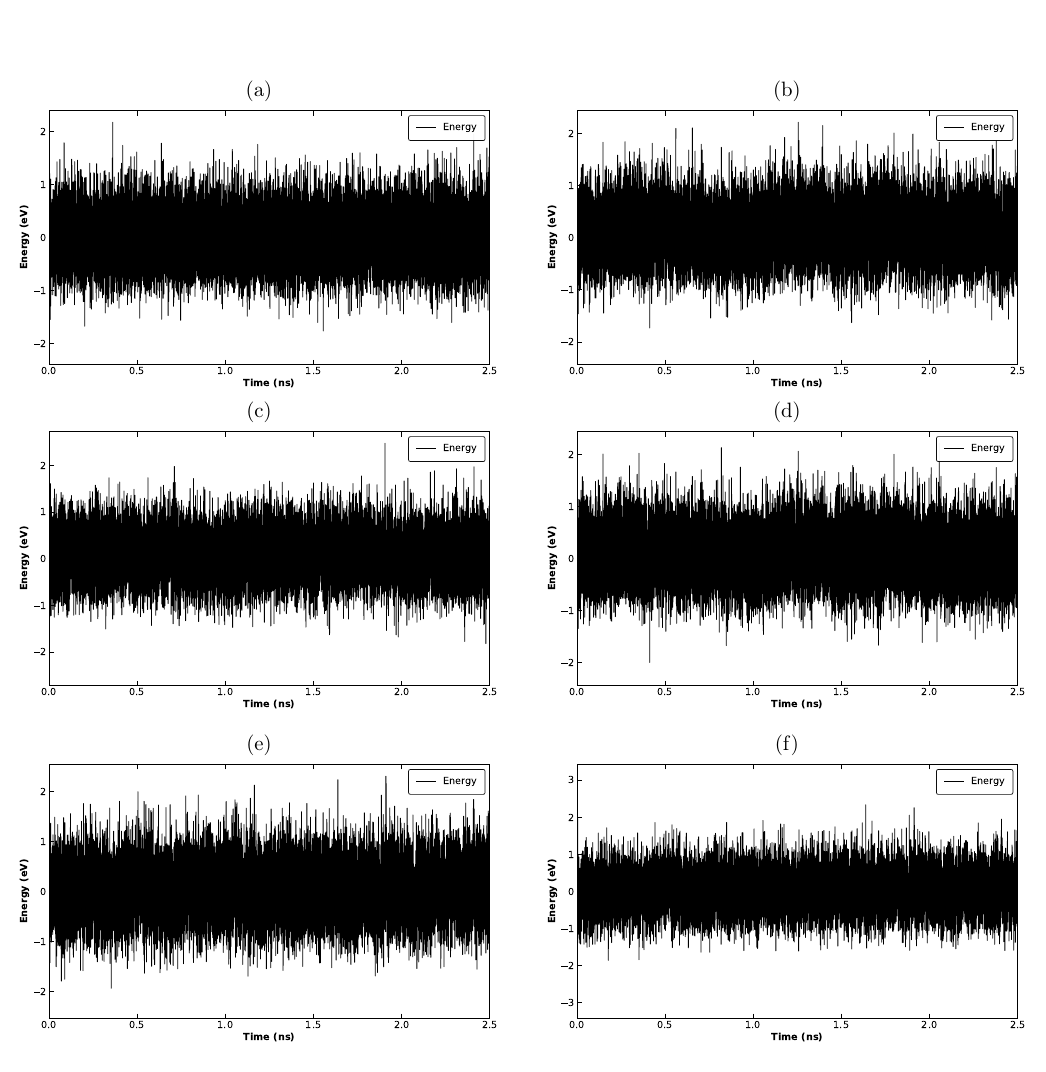}
\caption{Energy as a function of time during the 2.5 ns MLFF molecular-dynamics trajectories at 300 K for (a) MoS, (b) MoSe, (c) MoTe, (d) WS, (e) WSe, and (f) WTe. The stationary energy fluctuations show that the MLFF trajectories remain bounded over the sampled time interval. This finite-temperature test probes the near-equilibrium thermal domain and should not by itself be interpreted as validation of strongly bent configurations.}
\label{si:fig_energy_vs_time}
\end{figure}


\clearpage
\section{Additional bending configurations}

\begin{figure}[H]
\centering
\includegraphics[width=\linewidth,clip]{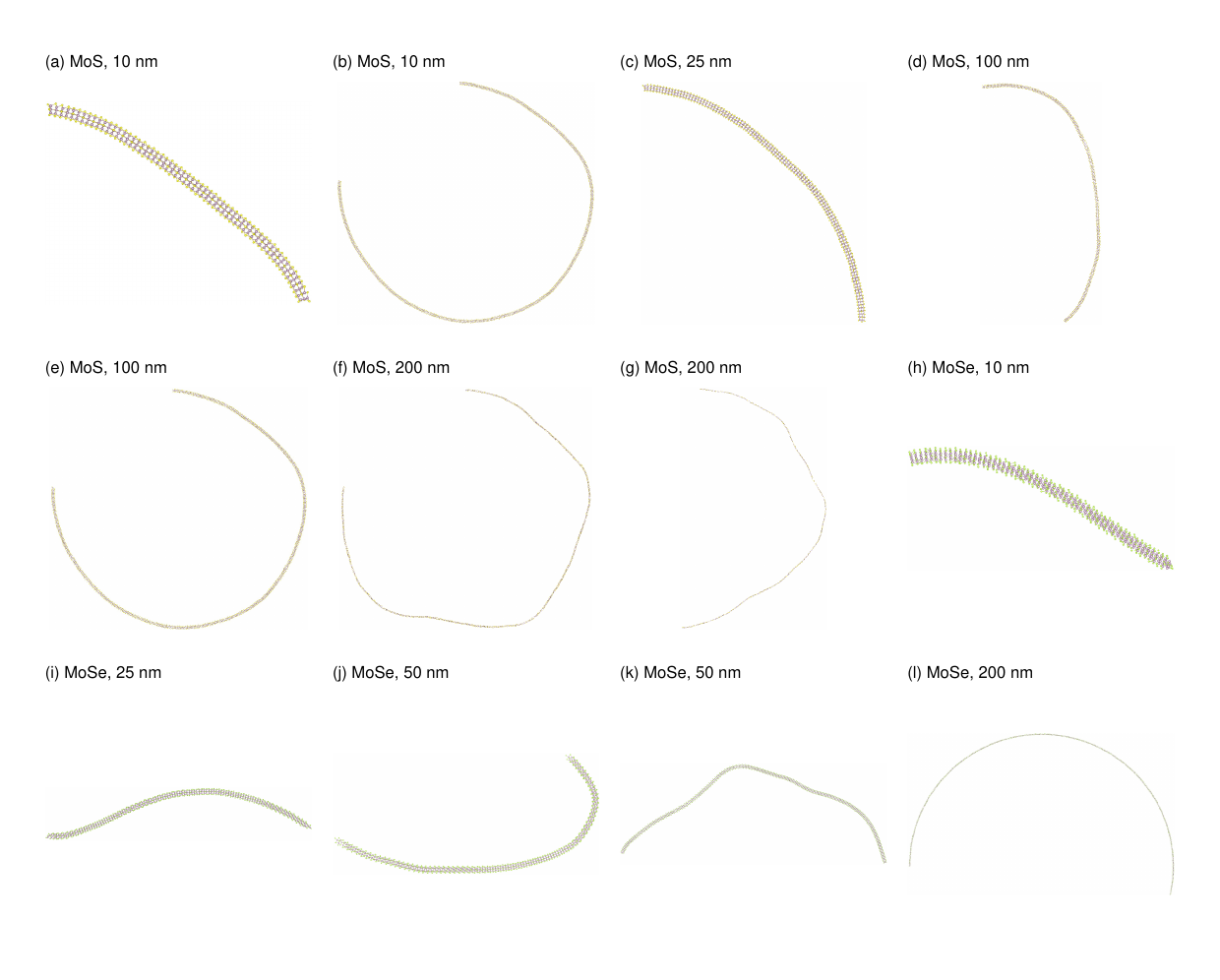}
\caption{Additional structural snapshots of bent MoS and MoSe nanowires.}
\label{si:bending_MoS_MoSe}
\end{figure}

\begin{figure}[H]
\centering
\includegraphics[width=\linewidth,clip]{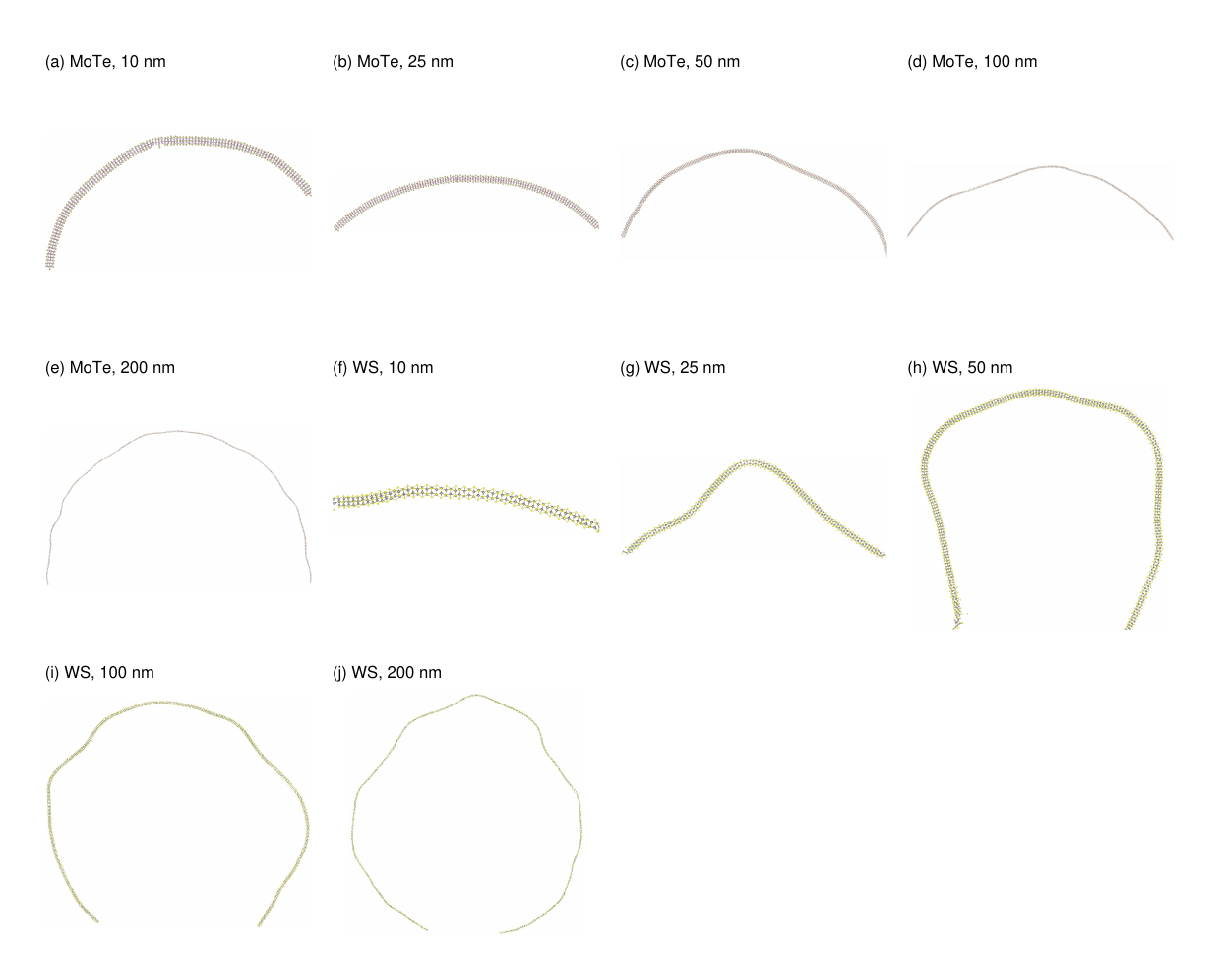}
\caption{Additional structural snapshots of bent MoTe and WS nanowires.}
\label{si:bending_MoTe_WS}
\end{figure}

\begin{figure}[H]
\centering
\includegraphics[width=\linewidth,clip]{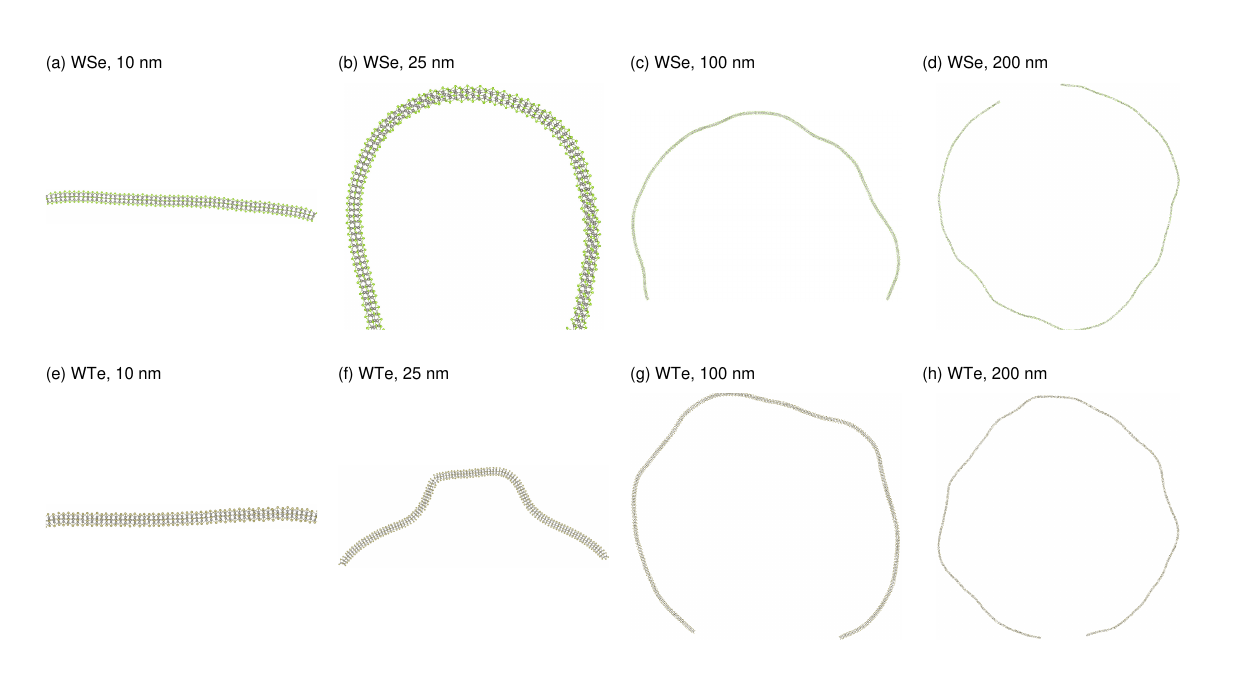}
\caption{Additional structural snapshots of bent WSe and WTe nanowires.}
\label{si:bending_WSe_WTe}
\end{figure}

\end{document}